\documentclass[preprint,showpacs,preprintnumbers,amsmath,amssymb]{revtex4}%
\usepackage{amsmath}
\usepackage{subfig}
\usepackage{graphicx}
\begin{document}

\title{ Instability in interacting  dark sector: An appropriate   Holographic Ricci dark energy model}

\author{Ram\'on Herrera}
\email{ramon.herrera@pucv.cl} \affiliation{Instituto de
F\'{\i}sica, Pontificia Universidad Cat\'{o}lica de
Valpara\'{\i}so, Avenida Brasil 2950, Casilla 4059,
Valpara\'{\i}so, Chile.}

\author{W. S. Hip\'olito-Ricaldi}
\email{wiliam.ricaldi@ufes.br} \affiliation{Departamento de
Ci\^encias Naturais,  Universidade Federal do Esp\'{\i}rito Santo, \\ Rodovia BR 101 Norte, km. 60,
S\~ao Mateus, Esp\'{\i}rito Santo, Brasil}

\affiliation{Department of Physics, McGill University, \\ Montr\'eal, QC, H3A 2T8, Canada.}
\author{Nelson Videla}
\email{nelson.videla@ing.uchile.cl} \affiliation{Departamento de
F\'{\i}sica, FCFM, Blanco Encalada 2008, Santiago, Universidad de
Chile, Chile.}

\date{\today}

\begin{abstract}

 In this paper we investigate the consequences of phantom
crossing considering  the perturbative dynamics in models with
interaction in their dark sector. By mean of  a general  study  of
gauge-invariant variables in comoving gauge, we relate  the
sources of instabilities in the structure formation process  with
the phantom crossing.  In order to illustrate  these  relations
and its consequences in more detail, we consider  a specific case
of an holographic dark energy interacting with dark matter.  We
find that in spite of the model is in excellent agreement  with
observational data at background level, however it is plagued of
instabilities in its perturbative dynamics.  We reconstruct the
model in order to avoid these undesirable instabilities, and 
we show that  this implies  a modification of the concordance
model at background. Also we find  drastic changes  on the
parameters space in our model when instabilities are avoided.

\end{abstract}
\pacs{98.80.Cq}
\maketitle

\section{Introduction}
It is well known that the measurements of the luminosity redshift
of supernovae (type Ia) have proportioned growing evidence for a
phase of accelerated expansion of current universe\cite{C1,C2}.
However,  other evidences of this accelerated expansion  come from
baryon acoustic oscillations \cite{C3}, anisotropies of the cosmic
microwave background (CMB)\cite{C4}, and among other \cite{C5},
have confirmed this scenario. In order to obtain a phase of
accelerated expansion in Einstein's General Relativity it is
necessary  that the cosmological background dynamics be dominated
by some exotic component with a negative pressure, known as dark
energy (DE).
 Assuming  that  DE contributes to an important   fraction of the content of the observable universe, it
 is instinctive and natural   from the  field theory to assume  its interactions with other fields, e.g.,
 dark matter (DM). It is well known that
  a suitable   interaction between DE and DM can provide an novel mechanism to
 alleviate the cosmic coincidence problem \cite{C6,C7}.  On the other hand, these models affect the structure formation and hence  provide
a different  way to change  the predictions of
 non-interacting models. Regarding this point,
the interaction between  DE and DM  have been studied  considering
different types of  observational data sets, see Refs.\cite{C8,C9}. For more comprehensive references
of models with interacting DE and DM, see Refs.\cite{I1,I2}.

The study of structure formation in models of DE and DM, through
the cosmological perturbations theory, plays a fundamental  role
when these models are confronted with the observations \cite{AC1}.
These models imprint a signature  on the CMB power spectrum
\cite{AC2,AC3} and also the space of parameters is modified\cite{AC3,AC4}
. For this reason the analysis of the cosmological perturbations is important and also need to be 
well-behaved. In
particular for interacting models, the background dynamics  with
adiabatic initial conditions and the perturbation theory
 were analyzed  in Ref.\cite{C10}. Here, the perturbative dynamics realizes  unstable growing modes.
A further analysis in  models with an interacting DE component
together  with a constant equation of state (EoS) $w$,
was considered in Ref. \cite{C11}. Here the authors found that
perturbations were unstable and with  a rapid growth of DE
fluctuations. To avoid a possible conflict with the perturbative
dynamics when the EoS parameter $w$ crosses the value $w=-1$, in
Ref.\cite{Kunz:2006wc}, the authors considered a new variable
associated to the divergence of the velocity field. However, the
new perturbations equations have a term associated to the pressure
perturbation and, therefore, an adiabatic speed of sound. In this
way, the authors introduced a free parameter in the adiabatic
speed of sound, avoiding the divergences in the perturbations.

On the other hand, considering that observational-data tests of the  $\Lambda$ cold dark matter ($\Lambda$CDM)
model are not accurate  enough to rule out adequately
 the large  diversity  of alternative DE models in the literature that have been proposed to
account for
the data. In modern cosmology, one  can
 test  the $\Lambda$CDM  to describe an adequate  DE model, assuming the DE  as an effective fluid (or  a scalar field)  and
considering its EoS  as  a free and dynamics parameter. Some candidates for this DE are
the
holographic models (HM) which give a specific classes of dynamic approaches to solve the cosmic coincidence problem
and  is another  alternative to the standard  $\Lambda$CDM model.
These models are motivated from the Holographic principle which  has its origin in  the black
hole and string theories \cite{HP1}.
 These HM have  a direct connection  between an ultraviolet and infrared
  cutoff \cite{HM1,HMM1,HMM2}. In this context,
  the infrared cutoff corresponds to a cosmological length scale, and by the other hand this connection
  between cutoffs ensures  that the energy density  does not exceed
  the energy in a given volume of a black hole of the equal size.
 Regarding the HM with interaction between DE and DM, this kind of models were studied in Refs.\cite{H0,HM2,HM3}.  In particular
  we mention
  a specific model of HM in which the cutoff length is proportional to the Ricci
  scale\cite{R1,V}, see also Ref.\cite{V1}. In relation to the study of the dynamics of
  perturbations, this was analyzed in Refs.\cite{V0,V2}. The appearance of instabilities in the dark sector, through the perturbative dynamics,
 occurs when the EoS parameter $w$ crosses the value $w=-1$ in models with a dynamical EoS.
   In this context,  the study of this crossing of the EoS parameter and the appearance of instabilities in the perturbative dynamics  for
   non-interacting  Ricci holographic model, was performed in Ref.\cite{DelCampo2013}.

 In the present paper we  study the background dynamics and
  also present the analysis of linear perturbations in the framework
of  gauge-invariant variables in comoving gauge, for the interacting dark sector,
 identifying the source of instabilities.
 In particular we consider that the dark energy density corresponds to the holographic Ricci DE model, and then we
 extended the study
 developed in Ref.\cite{DelCampo2013}, but now considering an interacting dark sector. Here we
 study the background equations, the linear perturbations and the appearance of instabilities.
 In order to evade these instabilities we develop an appropriate  holographic Ricci interacting  dark energy model
 and we find drastic changes on the constraints of the parameters.

 This article is organized as follows. In Sect. II we present the background
  equations for the interacting dark sector. In Sect. III we analyze the dynamics of perturbations in a general framework
  for a single fluid and two interacting fluids. In Sect. IV we study the evolution of linear
  perturbations and identify the sources of instabilities. In Sect. V we consider a
  specific holographic dark energy model, known as Ricci DE. Here we study the background dynamics and analyze the
  observational tests on this model, considering the SNIa and $H(z)$ data sets. In Sect.VI we study the linear
  perturbations and identify the instabilities in our interacting model. We also analyze the high-redshift limit of our model and we compare it with $\Lambda$CDM model. In Sect. VII we study how to avoid these instabilities and develop an appropriate model .
  Finally, Sect. VIII summarizes our results and exhibits our conclusions.

\section{Background equations:  dark energy- dark matter interaction }

We consider a spatially flat Friedmann- Roberson-Walker universe
dominated by two interacting  components, dark energy (subscript
$x$) and dark matter (subscript $m$) that behaves as
pressureless dust. In this form, the total energy density
is given by $\epsilon =\epsilon_m+\epsilon_x$ and the Friedmann equation can
be written as

\begin{eqnarray}
\label{Fr1} 3H^2=\epsilon=\epsilon_m+\epsilon_x,
\end{eqnarray}
where $H=\dot{a}/a$ is the Hubble rate and $a$ is the scale factor.
For convenience we will use the units in which $8\pi G=c=\hbar=1$,
and the dots mean derivatives with respect to the cosmological time.

On the other hand, we assume that both energy densities do
not evolve separately, but rather they interact with each other through
a source term, that enters the energy balance equations as
\begin{eqnarray}
\label{balance1} \dot{\epsilon}_m+3H\epsilon_m
=Q,\,\,\;\;\mbox{and}\,\,\,\;\;\;\dot{\epsilon}_x+3H(\epsilon_x+p_x)=-Q,
\end{eqnarray}
here $Q$ denotes the interaction term. If $Q>0$, the direction of energy transfer is
from DE to DM, if $Q$ is negative then the direction of energy transfer occurs from
DM to DE.
We note that the total
energy density $\epsilon=\epsilon_m+\epsilon_x$ is conserved. We
also consider that the DE component obeys an  EoS, such that
$w_x\equiv\frac{p_x}{\epsilon_x}=\frac{p}{\epsilon_x}\equiv w$,
where $w$ corresponds to  the EoS parameter. Here the quantity $p_x$ denotes  the pressure
associated with the DE. In virtue of these quantities,  the
acceleration equation becomes
\begin{eqnarray}
\label{Hdot} \dot{H} &=&
-\frac{3}{2}H^2\left(1+\frac{w}{1+r}\right),\,\,\;\mbox{or
equivalently} \;\,\,\,\,\,\frac{d \ln H}{d \ln a}
=-\frac{3}{2}\left(1+\frac{w(a)}{1+r(a)}\right),
\end{eqnarray}
where $r\equiv \frac{\epsilon_m}{\epsilon_x}$ denotes the ratio
between both energy densities. We also note that the \emph{total}
effective EoS of the cosmic medium $w_{total}=p/\epsilon$ can be
written as
\begin{eqnarray}
\label{total w}
w_{total}=\frac{p}{\epsilon}=\frac{p_x}{\epsilon_m+\epsilon_x}=\frac{w}{1+r}.
\end{eqnarray}
From (\ref{balance1}), the rate of change of the ratio between both
energy densities $r$ becomes
\begin{eqnarray}
\label{rdot}
\dot{r}=3Hr(1+r)\left[\frac{w}{1+r}+q_m\right]=3Hr(1+r)\left[w_{total}+q_m\right],
\end{eqnarray}
where the quantity $q_m$ is defined as
\begin{eqnarray}
 q_m\equiv
\frac{Q}{3H\epsilon_m}.\label{Q}
\end{eqnarray}
In particular in  the absence of interaction, we have that  $q_m=0$.

\section{General remarks on perturbations:
interacting  model}
\subsection{Single fluid}
In order to motivate the analysis of cosmological perturbations in DE models and its dynamics, we start by reviewing the
perturbations for a single fluid.

We consider the dark sector has an energy-momentum tensor of a
perfect fluid given by $T_{\mu\nu}=\epsilon u_\mu u_\nu+ p
h_{\mu\nu}$, where the tensor $h_{\mu\nu}$ is defined as
$h_{\mu\nu}=g_{\mu\nu}+u_{\mu}u_{\nu}$ and $u_\mu\,u^{\mu}=-1$.
Here, again $\epsilon$ represents the energy density, $p$ the
pressure and $u^{\mu}$ corresponds to  the 4-velocity of the dark
fluid. From the conservation law of the energy-momentum tensor
$T^{\mu\nu}_{;\nu}$=0, we obtain that the equations for timelike and spacelike
parts  are given by
\begin{eqnarray} \label{energybalancedarksector}
\epsilon_{,\alpha}u^\alpha+\Theta\left(\epsilon+p \right) = 0
\,,
\end{eqnarray}
and
\begin{eqnarray}
\left(\epsilon+p\right)\dot{u}^\mu+p_{,\alpha}h^{\alpha
\mu}=0 \,,\label{ecmom}
\end{eqnarray}
respectively. Here, the quantity $\Theta \equiv u^\mu_{;\mu}$ is the
expansion scalar and $\dot{u}\equiv u^{\mu}_{;\nu}u^{\nu}$.

In order to  calculate the perturbations, we consider the most general flat metric
, containing only scalar perturbations of a homogeneous and
isotropic background given by
\begin{eqnarray}
\mbox{d}s^{2} = - \left(1 + 2 \phi\right)\mbox{d}t^2 + 2 a^2
F_{,i}\mbox{d}t\mbox{d}x^{i} + a^2\left[\left(1-2\psi\right)\delta
_{ij} \right. + \left.2E_{,i\,j} \right] \mbox{d}x^i\mbox{d}x^j ,
\label{ds} \,
\end{eqnarray}
where $\phi=\phi(t,{\bf{x}})$, $F=F(t,{\bf{x}})$,
$\psi=\psi(t,{\bf{x}})$ and $E=E(t,{\bf{x}})$ are metric perturbation, see Ref.\cite{bardee}.

For the  4-velocity $u_{\mu}=(u_0,u_i)$ we get
\begin{eqnarray}
\label{tetrav}
 u_0 =1-\phi ,\;\;\;\;\;\mbox{and}\qquad \, u_i= v_{,i} \,,
\end{eqnarray}
where $v$ denotes  the scalar velocity perturbation.  Considering the metric (\ref{ds}), we find that the quantity
 $\Theta=3H$,  and its perturbation $\delta \Theta$
is given by
\begin{eqnarray}
\delta \Theta= \frac{\nabla^2 }{a^2}\left(v+  \chi\right)-3\dot{\psi}-3H\phi\,,\,\,\,\mbox{where}
 \qquad \chi=\left(\dot{E}-F\right) \,.
\end{eqnarray}
The Eq. (\ref{energybalancedarksector})  can be rewritten    up to  zeroth and also to first order
of perturbations, yielding
\begin{eqnarray} \label{darksector}
\dot{\epsilon}+\Theta\left(\epsilon+p\right)=0  \,,\,\,\,\mbox{and} \qquad
 \dot{\delta \epsilon}-\dot{\epsilon}\phi+\delta \Theta \left(\epsilon+p\right)+\Theta\left(\delta \epsilon+\delta p\right)=0 \,.
\end{eqnarray}
Usually the right equation of (\ref{darksector}) can be written in
function of  density contrast defined as $\delta=\frac{\delta
\epsilon}{\epsilon}$. However, the quantity $\delta$  is not
gauge-invariant. Thus is suitable  to describe the
dynamics of perturbations  in terms of gauge-invariant variables.
In particular, in  comoving gauge, these invariant quantities
represent perturbations on comoving hypersurfaces. In the
following,  we will denote a superscript $c$ to all
gauge-invariants in comoving gauge. These invariant quantities
are defined as $\delta^c=\delta
+\frac{\dot{\epsilon}}{\epsilon}\,v$, $\delta \Theta^c=\delta
\Theta+\dot{\Theta}v$ and $\delta p^c=\delta p + \dot{p}\,v$,  see
Ref.\cite{Hipolito2009}. In virtue of these gauge-invariant
quantities the right equation of (\ref{darksector}) can be
rewritten as
\begin{eqnarray}
\label{deltam}
\dot{\delta}^c-\Theta \frac{p}{\epsilon}\delta^c+\left(1+\frac{p}{\epsilon}\right)\delta
\Theta^c=0\,,
\end{eqnarray}
or equivalently
\begin{eqnarray}
\delta^{c\prime\prime} + \left[\frac{3}{2}-\frac{15}{2}\frac{p}{\epsilon}+
 3\frac{p^{\prime}}{\epsilon^{\prime}}\right]\frac{\delta^{c\prime}}{a}
- \left[\frac{3}{2} + 12\frac{p}{\epsilon} - \frac{9}{2}\frac{p^{2}}{\epsilon^{2}} - 9\frac{p^{\prime}}{\epsilon^{\prime}}
\right]\frac{\delta^{c}}{a^{2}}
+ \frac{k^{2}}{a^{2}H^{2}}\frac{\delta p^{c}}{\epsilon a^{2}}
= 0\,,
  \label{dddeltak}
\end{eqnarray}
where the primes denote derivatives with respect to the scale factor $a$ (for more
details, see Refs.\cite{vomMarttens2014,Funo2014}).
Here we point out that this equation governs the dynamics  of perturbations for the dark sector  as a whole.
Also considering  at first order  Eq.(\ref{ecmom}),  we  find  the momentum  balance equation
becomes
 \begin{eqnarray}
 \dot{v}+\phi+\frac{\delta p^c}{\epsilon+p}=0 \,.
\end{eqnarray}

\subsection{Interacting Two-component fluid: General Formalism }
In this subsection we consider a general formalism to study the perturbative dynamics for two interacting
fluids: DM that behaves as pressureless dust and DE. In the following, we will consider that both components are interacting and
then the energy momentum tensor for each individual component is not conserved separately, i.e.,
$T^{\mu\nu}_{A;\nu}=Q^\mu_A$. Here
 $Q^\mu_A$ denotes the  energy-momentum
transfer vector between both fluids  and "A'' label denotes both components: $A=m$ for dark matter and $A=x$ for dark energy.

 By considering the timelike part of the  balance equation and  from the projection in the direction of the vector
  $u_{A\mu}$, we find that
\begin{eqnarray} \label{energybalance}
\epsilon_{A,\mu}u^\mu_A+\Theta_A\left(\epsilon_A+p_A\right) = -u_{A \mu}Q^\mu_A = Q_A\,.
\end{eqnarray}
 In general  the expansion scalar $\Theta_A =u^\mu_{A;\mu}$
is  different for each component of the dark sector, however  up to zero order, or equivalently at the background level,
the 4-velocities are $u^{\mu}_m=u^{\mu}_x=u^{\mu}$, then $\Theta_A=\Theta =3H$. We emphasize
 that Eq.(\ref{energybalance})  corresponds to  the  projections  of vector $Q^{\alpha}_A$ along the 4-velocity
 $u_{A\alpha}$. In this way, the scalar quantity $u_{m\alpha}Q^{\alpha}=
- u_{x\alpha}Q^{\alpha}_x=u_{\alpha}Q^{\alpha}= Q$. On the order hand, the perturbed time components of the 4-velocities, 
to first order, are given by $\delta u_0=\delta u^0=\delta u^0_m=\delta u^0_x=-\phi$.

Now the energy-momentum balance equations are given by taking the spacelike part of  the vector
 $T^{\mu\nu}_{A;\nu}=Q^\mu_A$, resulting in
\begin{eqnarray} \label{momento}
 \left(\epsilon_A+p_A\right)\dot{u}_{A \mu}+p_{A,\alpha}h^{\alpha}_{A \mu} = h_{A\mu\alpha}Q^\alpha_A ={\cal{Q}}_{A \mu}
 \,,
\end{eqnarray}
where again $\dot{u}_{A,\beta}=u^{\alpha}_{A;\beta}u_{A \alpha}$ and $p_A$ denotes
 the pressure of the
 $A$-fluid. Following Refs.\cite{q1,q2}  the energy-momentum transfer
 $Q^\alpha$, can be   decomposed   in two parts, one  proportional
and other perpendicular to the total 4-velocity $u^\alpha$, so that
\begin{eqnarray}
Q^\alpha= u^{\alpha}Q+{\cal{Q}}^{\alpha}, \,\,\,\mbox{such that} \qquad u_\alpha {\cal{Q}}^{\alpha}=0 ,
\,\,\,\mbox{and} \qquad  Q=-u_{\mu}Q^{\mu}.
\end{eqnarray}

Also, we note that this decomposition of the vector $Q^\alpha$ implies that at first order,  ${\cal{Q}}^{\alpha}=(0,{\cal{Q}}^i)$, where
 ${\cal{Q}}^i$  corresponds to spatial vector up to  first order. In virtue of
 these quantities,
Eq. (\ref{energybalance}) for dark matter and for dark energy  may be rewritten as
\begin{eqnarray}
\epsilon_{m,\alpha}u^\alpha+\Theta_m\,\epsilon_m = Q  \,, \,\;\;\;\mbox{and}
\qquad \epsilon_{x,\alpha}u^\alpha+\Theta_x\,(\epsilon_x+p_x) = -Q \,,
\end{eqnarray}
respectively. For  the dark matter, we find that the
energy balance can be obtained considering   at first order of Eq.
(\ref{energybalance}) in which
\begin{eqnarray}
\label{matter}
\dot{\delta \epsilon}_m-\dot{\epsilon}_m\phi+\Theta \delta \epsilon_m + \epsilon_m \delta \Theta_m
= -\delta \left(u_{m\alpha} Q^{\alpha}\right) = \delta Q  \,,
\end{eqnarray}
whereas  for the dark energy we get
\begin{eqnarray}
\label{darkenergy}
\dot{\delta \epsilon}_x-\dot{\epsilon}_x\phi+\Theta \left(\delta \epsilon_x+\delta p_x\right) +
\delta \Theta_x\left(\epsilon_x+ p_x\right) = \delta \left(u_{x\alpha} Q^{\alpha}\right) =-\delta Q \,.
\end{eqnarray}
Here we have introduced  the  functions of contrast of matter
density $\delta_m =\frac{\delta \epsilon_m}{\epsilon_m}$ and dark
energy density $\delta_x =\frac{\delta \epsilon_x}{\epsilon_x}$
respectively. However, these densities  are not
gauge-invariants and  we shall describe our results in terms of
gauge-invariants in comoving gauge.  Following
Ref.\cite{Vargas2012} we will consider the
 invariant quantities $\delta \Theta^c_m = \delta \Theta_m+ \dot{\Theta}v$,
 $\delta^c_m=\delta _m+\frac{\dot{\epsilon}_m}{\epsilon_m} v$, and
$\delta Q^c = \delta Q+\dot{Q}v$ in the comoving gauge. In this way,  at  first order the
gauge-invariant equation for the dark matter contrast can be  rewritten as
\begin{eqnarray}\label{matter2}
 \dot{\delta}^c_m+\frac{\dot{\epsilon}_m}{\epsilon_m}\frac{\delta p^c}{\epsilon+p}+
 \delta \Theta^c_m =\frac{\delta
 Q^c}{\epsilon_m}-\frac{Q}{\epsilon_m}\delta^c_m.
\end{eqnarray}

Considering the most general perturbed metric given by Eq.(\ref{ds}),  we find that
the scalar perturbation
$\Theta_A$ can be written as  \cite{Funo2014}
\begin{eqnarray}
\label{thetaA}
\delta \Theta_A= \frac{\nabla^2}{a^2}\left(v_A+ \chi\right)-3\dot{\psi}-3H\phi\,, \,\,\,\,\mbox{where}\qquad
\chi=\left(\dot{E}-F\right).
\end{eqnarray}
Here we have considered that
  $\Theta_A=u^\mu_{A;\mu}$, which corresponds to  the scalar expansion for the $A$-component.

Also,  at first order  the energy-momentum balance Eq.(\ref{momento}) for both
dark fluids becomes
\begin{eqnarray} \label{a17}
 \epsilon_m\left(\dot{v}_m+\phi\right)_{,i}={\cal {Q}}_i \,,\,\,\,\mbox{and}\qquad \left(\epsilon_x+p_x\right)\left(\dot{v}_x+\phi\right)_{,i}
 + \delta p^{c_x}_{,i}=-{\cal {Q}}_i ,
\end{eqnarray}
respectively. Here we mention that the combination $\dot{v}+\phi$
is a gauge-invariant quantity in the comoving gauge, and the
quantity $p^{c_x}$  is defined as $p^{c_x}\equiv \delta
p_x^c+\dot{p}_x v_x=\delta p^c+\dot{p}v_x$ i.e.,
comoving to the dark energy

\section{Linear perturbations  and source of instabilities: Relative energy-density perturbation }
  In order to  analyze  the perturbative dynamics in terms of gauge-invariant quantities, we will
study the linear perturbations considering the relative energy-density perturbation $S^c$.  Following
Ref.\cite{DelCampo2013} we introduce the relative energy-density perturbation
$S^c$,
defined as $S^c= \Delta^c-\delta^c_m$, where $\Delta^c= \frac{\delta \epsilon^c}{\epsilon+p}$.
As it was noticed in Ref.\cite{DelCampo2013}, the instabilities in the perturbative dynamics of the fluids
are described in terms of this function.

 In order to find the equation for the  variable $S^c$,  we need to rewrite Eq.(\ref{deltam}) in
  terms of the dimensionless variable $\Delta^c$. Considering
Ref.\cite{Hipolito2009} we have that
\begin{eqnarray}
\label{Density}
 \Delta^c-\Theta \frac{\dot{p}}{\dot{\epsilon}}\Delta^c+\delta \Theta^c=0 \,,
\end{eqnarray}
and now, combining   with  Eq.(\ref{matter2}), we find that the equation for the relative perturbation $S^c$ results
\begin{eqnarray}
\label{eseQ}
 \dot{S}^c+\Theta\,\frac{\delta p^c_{nad}}{\epsilon+p}+\delta \Theta^c -
 \delta \Theta^c_m=G \,,
 \end{eqnarray}
 where the function $G$ is defined as
 $$
 G=G(q_m,\delta Q^c) = -\frac{\delta Q^c}{\epsilon_m}+3Hq_m\delta^c_m+3Hq_m\frac{\delta
 p^c}{\epsilon+p}.
 $$
 Here, $\delta p^c$ corresponds to  the gauge-invariant expression for pressure perturbation of the dark fluid,
 and the quantity  $\delta p^c_{nad}$ denotes  the non-adiabatic contribution of the pressure of the dark
 fluid.
 We mention that the relation between both quantities is given by
  $\delta p_{nad}= \delta p^c-\frac{\dot{p}}{\dot{\epsilon}}\delta \epsilon^c$. Also, by considering
the interaction to first order and  the comoving gauge, the quantities $\delta Q^c$
and $\delta \Theta^c$ are defined as
 $\delta Q^c =\delta Q+\dot{Q}v$ and
 $\delta \Theta^c=\delta \Theta+\dot{\Theta}v$, respectively.

Adding   Eqs.(\ref{matter}) and (\ref{darkenergy})  and
comparing with Eq.(\ref{darksector}),
 we get
\begin{eqnarray}
 \delta \Theta^c =\frac{\epsilon_m}{\epsilon+p}\delta \Theta_m^c+ \frac{\epsilon_x+p_x}{\epsilon+p}\delta \Theta_x^c \,,
\end{eqnarray}
which allow us to write
\begin{eqnarray}
\label{thetadif}
\delta \Theta^c - \delta \Theta_m^c =\left(1-\frac{\epsilon_m}{\epsilon+p}\right)\left(\delta \Theta_x^c-\delta \Theta_m^c\right)=
 \left(1-\frac{\epsilon_m}{\epsilon+p}\right)\frac{\nabla^2}{a^2}\left(v_x-v_m\right) \,,
\end{eqnarray}
where we have used Eq.(\ref{thetaA}).  Now combining  Eqs.(\ref{a17}),
(\ref{eseQ}),
and (\ref{thetadif}), we  find that the equation for the relative energy-density becomes
\begin{eqnarray}
\label{SaaG}
S^{c \prime \prime }+\left(1+a\frac{H'}{H}+\tilde{A}(a)\right)\frac{S^{c \prime}}{a}+\tilde{B}(a)=0,
\end{eqnarray}
where
\begin{eqnarray}
\tilde{A}(a)=\frac{3r}{1+w}(q_m+\hat{c^2_s})+2 ,\,\qquad \hat{c}^2_s \equiv \frac{p'}{\epsilon
'},\label{AA}
\end{eqnarray}
and
$$
 \tilde{B}(a)= \frac{1}{a
H}\left[\frac{3H}{\epsilon+p}\delta p^c_{nad}\right]'+
\frac{\tilde{A}(a)}{a^2 H}\left[\frac{3H}{\epsilon+p}\delta
p^c_{nad}\right]-\frac{\nabla^2}{a^4 H^2}\frac{\delta
p^{c_x}}{\epsilon+p}
$$
\begin{eqnarray}
-\frac{{\cal{Q}}}{a^2H^2\epsilon_m}
 -\tilde{A}(a)\frac{G}{a^2 H}-\frac{G'}{aH}.\label{BQ}
\end{eqnarray}
Here,  we
observe that in the limit $Q \rightarrow 0$,  the function
$\tilde{A}(a)$ corresponds to the obtained in
Ref.\cite{DelCampo2013}. As before the primes denote derivatives
with respect to the scale factor and $\hat{c}_s$ denotes the total
 adiabatic speed of sound.

In this form, Eqs.(\ref{dddeltak}) and (\ref{SaaG})
 are the fundamental  equations that govern the dynamics of perturbations in the
case of interacting  DE and DM fluids,  since it allows us to find
the perturbations
 $\delta^c_m$ and $\delta^c_x$. We mention that
the Eqs.(\ref{dddeltak}) and (\ref{SaaG}) are  coupled (as we
shall see in the next sections) and the sources of this coupling
are the functions  $\delta p_{nad}$, $\delta p^c_x$, and the
interaction term  $q_m$.

At this point we observe that, if any DE model has
 a dynamic EoS and it crosses the value $w=-1$ in any finite time,
then there will exist a source of instabilities driven by  the
function $\tilde{A}(a)$. In particular, these instabilities shall
appear in matter perturbations via the $S^c$-function
independently of whether DE and DM are interacting or not.  As we
will mention  in the next subsection,  terms related to
$\delta p^c_{nad}$ and $G$ do not present divergences. However we
will have another contribution to the instabilities arising from
the term related to the pressure perturbations $\delta p^{c_x}$.

\subsection{Pressure terms and perturbative dynamics}
An interesting feature of the equations for $\delta^c$ and $S^c$
is that they  are directly related to the non-adiabatic total
pressure perturbation $\delta p ^c$ and the  dark energy
pressure perturbation $\delta p^c_x$. In the following,  we will
express these two pressure perturbations as functions of
$\delta^c$ and $S^c$, and then we will analyze the instabilities and its
sources. Following Ref.\cite{Hipolito2010}, the non-adiabatic part
of the total pressure perturbation $\delta p_{nad}$, can be written
as
\begin{eqnarray}\label{general na}
\delta p_{nad}^c= \delta p_{x,nad}^c + \epsilon_m
\frac{\epsilon_x+p_x}{\epsilon+p}\frac{\dot{p}_x}{\dot{\epsilon}_x}
\left(\frac{\delta
\epsilon_x^c}{\epsilon_x+p_x}-\delta_m^c\right),
\end{eqnarray}
where
$$
\,\delta p_{x,nad}^c=\delta
p_x^c-\frac{\dot{p}_x}{\dot{\epsilon}_x}\delta \epsilon_x^c \,.
$$

Here, $\delta p^c_{x,nad}$ corresponds to the non-adiabatic part of
the dark energy pressure perturbation,  which is intrinsic to
dark energy. The equation for the total pressure perturbation
$\delta p_{nad}$, given by Eq.(\ref{general na}), can be rewritten
as
\begin{eqnarray}
\delta p_{nad}^c= \delta
p_{x,nad}^c+c^2_s\frac{\epsilon_m}{\epsilon+p}\left[-q_m\epsilon\delta^c
+ (\epsilon+p)S^c\right],\label{os}
\end{eqnarray}
where the sound speed becomes
$$
c^2_s\equiv
\frac{p'_x}{\epsilon'_x}=w-\frac{1}{3}\frac{r(w+q_m+q_mr)}{q_mr+1+w}.
$$
 Here, we observe that in the limit $Q\rightarrow0$ the
sound speed $c^2_s$ has a divergence in the case when the EoS
parameter $w\rightarrow-1$ and the dynamic of the pressure
perturbations collapse. However, in the case with interaction, we
note that the sound speed is finite when $w \rightarrow -1$ and
this divergence does not occur in our case.

From Eq.(\ref{os}) we observe that the non-adiabaticity of the
dark sector arises from the non-adiabaticity of the dark energy
and  the relative entropy  (the second term of Eq.(\ref{general
na})) between  dark energy and dark matter fluid   and
also from the  interaction $q_m\propto Q$ (see
Eq.(\ref{Q})). Now, and considering for simplicity that
the dark energy is an adiabatic fluid, i.e., $\delta p_{x,nad}=0$,
then  from Eq.(\ref{os}) the contribution to the non-adiabatic
total pressure perturbation arises from the relative entropy
between DM-DE and the interaction term  $q_m$. In this form, the
total pressure perturbation $\delta p^c=\delta
p_{nad}+\dot{p}/\dot{\epsilon}\,\delta\epsilon^c$ reduces to
\begin{eqnarray}
\label{dpct1} \delta p^c=  (\hat{c}^2_s-q_m y_1)\epsilon \delta^c
+ c^2_s \epsilon \frac{r}{1+r} S^c \,,
\end{eqnarray}
where
\begin{eqnarray}
\hat{c}^2_s=\frac{c^2_s(q_mr+1+w)}{1+r+w} \,,\,\,\,\,\mbox{and}
\qquad y_1=\frac{rc^2_s}{1+r+w}.
\end{eqnarray}
From Eq.(\ref{dpct1})  the term $3H\,\frac{\delta
p_{nad}^c}{\epsilon+p}$ can be written as
\begin{eqnarray}
\label{dpnad}
3H\frac{\delta p_{nad}^c}{\epsilon+p}=f_1\delta^c + f_2 S^c,
\end{eqnarray}
where
\begin{eqnarray}
\label{dpnadfs}
f_1\equiv -3Hq_m(1+r)y_1,\,\,\,\,\,f_2=3Hy_1.
\end{eqnarray}

Under the assumption that $\delta p_{x,nad}=0$, the function
$G$ given by Eq.(\ref{eseQ}) takes the form
\begin{eqnarray}
\label{GQ}
G=-\frac{\delta Q^c}{\epsilon_m}+g_1 \delta^c+g_2 S^c,
\end{eqnarray}
where the coefficients $g_1$ and $g_2$ are given by
\begin{eqnarray}
\label{G1G2} g_1\equiv 3Hv_1(1+\hat{c}_s^2-q_m
y_1),\,\,\,\,\,g_2\equiv
3Hq_m\left(y_1-1\right),\,\,\;\;\mbox{and}\,\,\,\,\;\,\,v_1\equiv
\frac{q_m(1+r)}{1+r+w},
\end{eqnarray}
respectively. Here we have considered  Eq.(\ref{dpct1}).

In order to find the coupled set of equations for $\delta ^c$
and $S^c$, with a general interaction term $Q$ and thus $\delta
Q^c$,  we need to find an expression for the pressure perturbation
associated to the dark energy component  $\delta p^{c_x}$ such that
\begin{eqnarray}
\label{dpcx1} \delta p^{c_x}=\delta p^c+\dot{p}v_x=\delta
p^c+\dot{p}v+\dot{p}(v_x-v)=\delta p+\dot{p}(v_x-v),
\end{eqnarray}
where the difference between the scalar velocity perturbations 
$v_x-v$ is given by
\begin{eqnarray}
\label{defvxv} v_x-v=\frac{\epsilon_m}{\epsilon+p}(v_x-v_m).
\end{eqnarray}

Now, and going to the $k$-space, where $k$ denotes the magnitude of the
physical momentum ($k=|{\bf{k}}|$),  we find  that  the
difference between the velocity perturbation of DE and DM, $v_x-v_m$ results
\begin{eqnarray}
\label{difvxvk}
v_x-v_m=\frac{a^2}{k^2}\frac{\epsilon+p}{\epsilon_x+p_x}
\left[S^{c\prime}aH+(f_1-g_1)\delta^c+(f_2-g_2)S^c+\frac{\delta Q^c}{\epsilon_m}\right].
\end{eqnarray}
Here, we have considered Eqs.(\ref{eseQ}), (\ref{thetadif}), and (\ref{GQ}),
respectively.

In this way, the quantity $\frac{\delta p^{c_x}}{\epsilon+p}$
appearing in Eq.(\ref{BQ}) becomes
\begin{eqnarray}
\label{deltapcxQ}
\frac{\delta p^{c_x}}{\epsilon+p}&=&-3Hy_1\frac{k^2}{a^2}\left(1+q_m\frac{r}{1+w}\right)\left[(f_1-g_1)\delta^c+(f_2-g_2)S^c
+aHS^{c \prime }+\frac{\delta Q^c}{\epsilon_m}\right]\nonumber\\
&& +v_1\left(\frac{\hat{c}_s^2}{q_m}-y_1\right)\delta^c+y_1S^c.
\end{eqnarray}
We  note that  the expression given by Eq.(\ref{deltapcxQ}) which appears  in the Eq.(\ref{BQ}), is also (joint with $\tilde{A}(a)$) responsible of the instabilities
in the perturbative dynamics when the dynamical EoS parameter $w$ crosses the value $w=-1$.

Considering Eq.(\ref{dpct1}), we determine the equation for the
gauge-invariant quantity $\delta^c$  given by
$$
\delta^{c\,\prime\prime} +
\left[\frac{3}{2}-\frac{15}{2}\frac{w}{1+r}+
3\hat{c}^2_s\right]\frac{\delta^{c\,\prime}}{a} -
\left[\frac{3}{2} + \frac{12w}{1+r} -
\frac{9}{2}\left(\frac{w}{1+r}\right)^2 - 9\hat{c}^2_s
+\frac{k^2}{a^2H^2}(q_my_1-\hat{c}^2_s)
\right]\frac{\delta^{c}}{a^{2}}
$$
\begin{eqnarray}
 =-\frac{k^{2}c^2_s}{a^{2}H^{2}}\frac{r}{1+r}\frac{S^{c}}{a^{2}}\,.
  \label{deltafinal1}
\end{eqnarray}
Also from  Eqs.(\ref{SaaG}),  (\ref{BQ}), (\ref{dpnad}),
(\ref{GQ}), and (\ref{deltapcxQ}), the equation for the   relative
energy-density perturbation can be written
$$
 S^{c \,\prime
\prime}+\left[1+m+\frac{y_2}{H}\right]\frac{S^{c\,\prime}}{a}+\left[\frac{a}{H}y_2^{\,\prime}+
\frac{y_2}{H}\left(m+\frac{3}{2}+\frac{3}{2}\frac{w}{1+r}\right)+\frac{k^2y_1}{a^2H^2}\right]
\frac{S^c}{a^2}+\frac{v_2}{H}\frac{\delta^{c\,\prime}}{a}
$$
$$
+\left[\frac{a}{H}v_2^{\,\prime}+\frac{v_2}{H}\left(m+\frac{3}{2}+\frac{3}{2}\frac{w}{1+r}\right)+
 \frac{k^2v_1}{a^2H^2}\left(\frac{\hat{c}^2_s}{q_m}-y_1\right)\right]\frac{\delta^c}{a^2}
$$
\begin{eqnarray}
 =-\frac{1}{a^2H}\left(m+\frac{3}{2}+\frac{3}{2}\frac{w}{1+r}\right)\frac{\delta Q^c}
 {\epsilon_m}-\frac{1}{aH}\left(\frac{\delta Q^c}{\epsilon_m}\right)^{\,\prime},\label{SppQ}
\end{eqnarray}

where the function $m$ is defined as
\begin{eqnarray}
m=\frac{1}{2}-\frac{3}{2}\frac{w}{1+r}+\frac{3r}{1+w}(q_m+\hat{c}^2_s-q_m
y_1)-3y_1,
\end{eqnarray}
with
\begin{eqnarray}
 y_2=3H\left[y_1(1-q_m)+q_m\right] \,,\,\,\,\,\mbox{and} \qquad
 v_2=3Hv_1\left(q_my_1-c_s^2r-1-\hat{c}^2_s\right).
\end{eqnarray}

The general Eqs. (\ref{deltafinal1}) and (\ref{SppQ})
allow us to obtain the solutions for the
perturbations of the dark sector, consisting in DE and DM fluids, interacting through a
$Q$ term and, accordingly, its perturbation $\delta Q^c$. Also we note that in
the limit in which $Q$  equals to zero, the  Eqs. (\ref{deltafinal1}) and (\ref{SppQ})
reduce to the equations obtained in Ref.\cite{DelCampo2013}.

As a concrete example, in the next section we will study a
particular dynamical interacting dark energy model, where the
energy density of the dark energy component has a holographic
nature. In this form we will extend the work performed in
Ref.\cite{DelCampo2013} adding  an interaction between DM and
Ricci-DE fluids. Also we shall illustrate, how a
well-situated model from point of view of observational
background  tests, could be plagued of instabilities in its
perturbative dynamics.

\section{Ricci dark energy:  background dynamics}
In this section we describe a cosmological
interacting dark energy model, where the DE corresponds to the
 holographic Ricci model.

We begin by summarizing  the characteristic of holographic DE with an
energy density of energy  $\epsilon_{x=H}\equiv\epsilon_H$, which interacts
with dark matter $\epsilon_m$. Also the energy density of each component is related to the total
dark sector with energy density ($\epsilon=\epsilon_H+\epsilon_m$),
through $\epsilon_m =r(1+r)^{-1}\epsilon $ and
$\epsilon_H=(1+r)^{-1}\epsilon$, where, as before, the rate $r$ is
defined as $r=\frac{\epsilon_m}{\epsilon_H}$.

Following Refs.\cite{HM1,HMM1}  holographic energy density is given by
 \begin{equation}
\epsilon_H=3\,c^2 \,L^{-2},\label{hol}
\end{equation}
where $c^2$ is a constant  and the factor 3 was introduced for
mathematical convenience. On the other hand, the quantity $L$ is an infrared
cutoff scale.  Different alternatives of the infrared cutoff
$L$ have been studied in the literature, see e.g., Refs.\cite{HM1,R1,V}.

Differentiating Eq.(\ref{hol}) and considering Eq.(\ref{balance1})
we get
\begin{eqnarray}
 \frac{Q}{\epsilon_H}=2\frac{\dot{L}}{L}-3H(1+w)\,.
\end{eqnarray}
In particular for the special case in which $Q=0$, the EoS
parameter becomes $w=(2\dot{L}-3HL)/3HL$ and coincides with the obtained in Ref.\cite{DelCampo2011}.

 Following Refs.\cite{R1,V} the  cutoff scale is given by $L^2 = 6 R^{-1}$, where $R$
 corresponds to the
Ricci scalar $R=6(2H^2+\dot{H})$ and then, the dark energy density becomes
$\epsilon_H =3c^2(2H^2+\dot{H})$.

From Eqs.(\ref{Fr1}) and (\ref{Hdot}) the EoS parameter becomes

\begin{eqnarray}\label{rew}
w = \frac{1+r}{3} -\frac{2}{3c^2}, \,\,\,\,\,\mbox{then} \qquad
3\dot{w}=\dot{r},\,\,\;\;\mbox{and} \qquad r=r_0+3(w-w_0) \,,
\end{eqnarray}
here we observe that the parameter $c^2$ is associated to $r_0$
and $w_0$, such that $c^2=2(r_0-3w_0+1)^{-1}$.


Deriving the Ricci DE, for which $\epsilon_H\propto R$, and
combining with Eq.(\ref{balance1}), we find a  relation between
the EoS parameter $w$ and the interaction term $Q$, given by
\begin{eqnarray}\label{rw}
\frac{Q}{H\epsilon_H}=-\frac{3}{1+r}\left[rw-\frac{\dot{w}}{H}\right]\,.
\end{eqnarray}
In particular, in the non-interacting case, the relation
Eq.(\ref{rw}) becomes  $r=\dot{w}/(wH)$, and  the background  dynamics and its cosmological consequences  were studied
in Ref.\cite{DelCampo2013}.

\subsection{Specifying $Q$}
Since both dark components are assumed to interact with each other
through the term $Q$, we must to specify the energy transfer rate
$Q$ in order to find solutions for the model studied here. Several
possibilities have been studied in the literature for the transfer
rate $Q$, see e.g. Ref. \cite{H0, Int1}. The most commonly used term $Q$ depends
on the energy densities $\epsilon_m$, $\epsilon_x$,
$\epsilon=\epsilon_m+\epsilon_x$ or combinations these, multiplied
by a term with units of  the inverse of time, i.e., a rate, where
the rate is proportional to the Hubble parameter i.e.,
$Q=Q(H\epsilon_m,H\epsilon_x, H\epsilon)$.  Other type of the
energy transfer rate   was considered from reheating models where
this rate is just a constant \cite{Re} and also for the curvaton
field case\cite{q2} .

In the following we will consider that the transfer rate $Q$ is proportional to the Hubble
rate. In this form, we  consider
 an interaction term given by
\begin{eqnarray}
\label{interaction}
 Q=\frac{\Theta}{\epsilon}\left(\beta_1 \epsilon^2_m+\beta_2 \epsilon_m\epsilon_H\right)\,,
 \;\;\;\;\mbox{with}
 \qquad {\cal{Q}}_i=0\,,
\end{eqnarray}
where at background level $\Theta=3H$. Here the parameters $\beta_1$ and $\beta_2$
are constants.

We note that from the ansatz given by  Eq.(\ref{interaction}) we have
  four different alternatives arising, namely; the case $\beta_1=\beta_2=0$ agrees to the non-interacting case, the case $\beta_1=\beta_2=\beta$ that corresponds to the interaction
$Q=3H\epsilon_m$, the case  $\beta_1=0$ gives
$Q=3H\frac{\epsilon_m\epsilon_H}{\epsilon}$, and finally, the case
where $\beta_2=0$, which corresponds to
$Q=3H\frac{\epsilon_m^2}{\epsilon}$. These sets of energy transfer
rate  and others were analyzed and discussed in Ref.\cite{Int2}.

By considering the interaction term given by Eq. (\ref{interaction}), we find an
expression for the quantity $q_m$ in terms of the ratio
$r=\epsilon_m/\epsilon_H$, given by
\begin{eqnarray}\label{qm}
 q_m=\frac{Q}{3H\epsilon_m}=\left(\beta_1 \frac{r}{1+r}+\beta_2 \frac{1}{1+r}\right) \,.
\end{eqnarray}
 In order to achieve $q_m>0$ (or equivalently $Q>0$) and,
 considering that the
 rate $r$ satisfies the condition $r>0$, then the allowed range for the ratio $r$ becomes
 $0<r<-\beta_2/\beta_1$. Here, we note that one of the coupling
 constants should be negative.

 Now, by combining   Eqs. (\ref{rdot}), (\ref{rew}) and (\ref{qm})   we find
analytical solutions for the ratio $r=r(a)$ and the EoS parameter $w$ as functions of
the scale factor $a$, wherewith
\begin{eqnarray}
\label{rwQ}
r =\frac{r_0 D }{\left(Cr_0+D\right)a^{-3D} -Cr_0} \,, w =
\frac{D r_0-Cr_0\left(3w_0-r_0\right)+\left(3w_0-r_0\right)\left(Cr_0+D\right)a^{-3D}}{3[\left(Cr_0+D\right)a^{-3D}
-Cr_0]},
\end{eqnarray}
where the constants $C$ and $D$ are defined as
\begin{eqnarray}
\label{CDQ}
C=\beta_1+\frac{1}{3}\,, \,\,\,\,\,\mbox{and} \qquad D= \beta_2+w_0-\frac{r_0}{3}
\,,
\end{eqnarray}
respectively.

From Eq.(\ref{Hdot}) the Hubble rate as function of the scale factor can be
written as
\begin{eqnarray}
\label{friedmann}
H(a)=H_0 a^{-3/2}e^{\frac{3}{2}\int^1_a \frac{w}{1+r}d ln\,a'} \,,
\end{eqnarray}
where the rate $w/(1+r)$ that appears in the integral  is given by
\begin{eqnarray} \label{w1r}
\frac{w}{1+r}=\frac{1}{3}\frac{D r_0+\left(3w_0-r_0\right)[\left(Cr_0+D\right)a^{-3D}-Cr_0]}{Dr_0+
\left(Cr_0+D\right)a^{-3D}
-Cr_0}.
\end{eqnarray}
Here, we have considered that $H(a=a_0=1)=H_0$.

Now we shall
 perform an observational analysis of the model using the
most recent compilations of SNIa (JLA compilation  \cite{Betoule2014}) and $H(z)$ (\cite{Farooq2013}) data, only by considering background dynamics. In
 order to achieve this analysis and find the best-fit for the parameters which characterize our model,
 we have considered Eqs.(\ref{friedmann})  and (\ref{w1r}).

.

\subsection{Tests using SNIa and $H(z)$}
In order to develop the observational analysis of the background dynamics
we consider the SNIa test, using  the  JLA compilation  \cite{Betoule2014} with 740 data points and we also consider the  observational test
corresponding to  the recently updated Hubble $H(z)$ data \cite{Farooq2013}.
Our  tests are based on $\chi^2$-statistics, which will allow us to explore
the space of parameters only by considering the background dynamics.

 In our statistical analysis  we consider the function $\chi^2$ defined as
\begin{equation}
\chi^{2}(\theta)=\Delta y (\theta)^T\mathbf{C}^{-1} \Delta y(\theta)\,,
\label{chi2}
\end{equation}
where $\Delta y(\theta) = y_i-y(x_i;\theta)$, $\theta$ corresponds to  the free parameters, $\mathbf{C}$ denotes the covariance matrix of data $y_i$ and
$y(x_{i}\vert\theta)$  represents   the theoretical predictions
for a given set of parameters. In the space of parameters the best fit is found by minimizing the $\chi^2$-function and
the minimum of this function gives us an indication of the quality of the fit.

At this point, we consider  the tests associated to   distance modulus of type Ia supernovae,
which is defined by
\begin{eqnarray}
 \mu(z,\theta)=5\,log_{10}(d_L(z,\theta))+42.38-5\,log_{10} h \,,
\end{eqnarray}
where the luminosity distance is defined as
\begin{eqnarray}
 d_L(z,\theta)=(1+z)H_0\int^{z}_0\frac{dz'}{H(z',\theta)} \,,
\end{eqnarray}
and  $H_0 = 100 \, h \mathrm{km s^{-1} Mpc^{-1}}$. Here, the Hubble rate $H(z)$ in terms of the redshift $z$
 is given by Eq.(\ref{friedmann}), in which  $a = (1+z)^{-1}$. Observational data points of the luminosity-distance  modulus
 $\mu_{obs}$ were calculated
using the relation \cite{Betoule2014}
\begin{equation} \label{muobs}
\mu_{obs} = m^*_B-(M_B-\alpha X_1+\beta C)\,,
\end{equation}
where $m^*_B$ corresponds to the observed peak magnitude in rest
frame $B$ band and  the quantities $\alpha$, $\beta$ and $M_B$ are
nuisance parameters. On the other hand, the parameter  $X_1$ is related to the
time stretching of the light-curves, and $C$ corrects the color at
maximum brightness. In order to calculate completely the quantity
$\mu_{obs}$ and its covariance matrix we consider the method
suggested in \cite{Betoule2014} and the JLA compilation \cite{JLA}.
Moreover,  we perform an analysis by using  the compilation of the
recently updated $H(z)$ data \cite{Farooq2013}, which were derived
using the differential evolution of passively evolving  galaxies
as cosmic chronometers \cite{Simon2005,Stern2010}.  For a
combination of both tests we use the total  $\chi^2$   such
that
 $\chi^2_{total}=\chi^2_{SNIa}+\chi^2_{H(z)}$.

\begin{figure}[htb]
\centering
  \begin{tabular}{@{}cc@{}}
    \includegraphics[width=.35\textwidth]{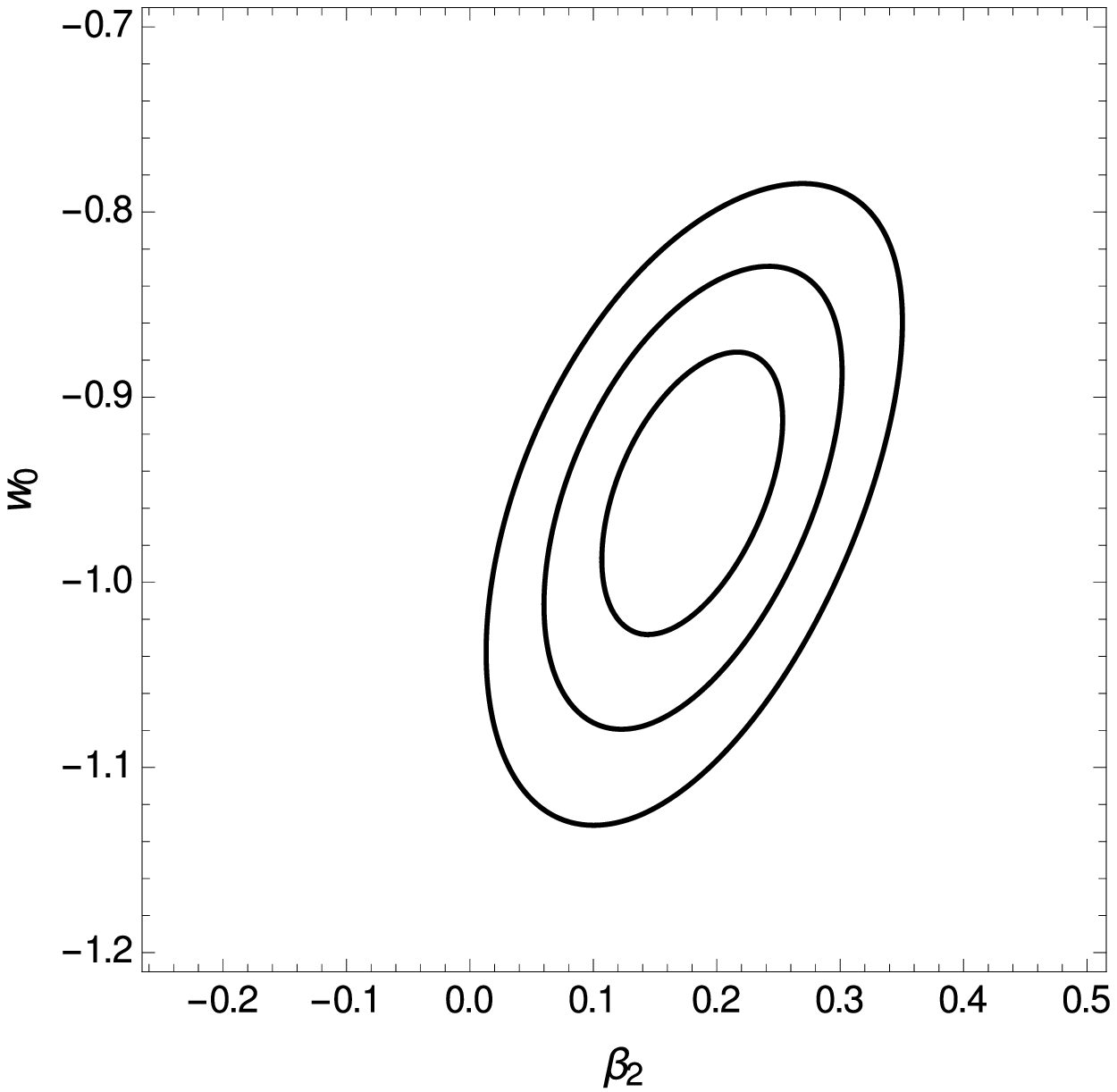} &
    \includegraphics[width=.35\textwidth]{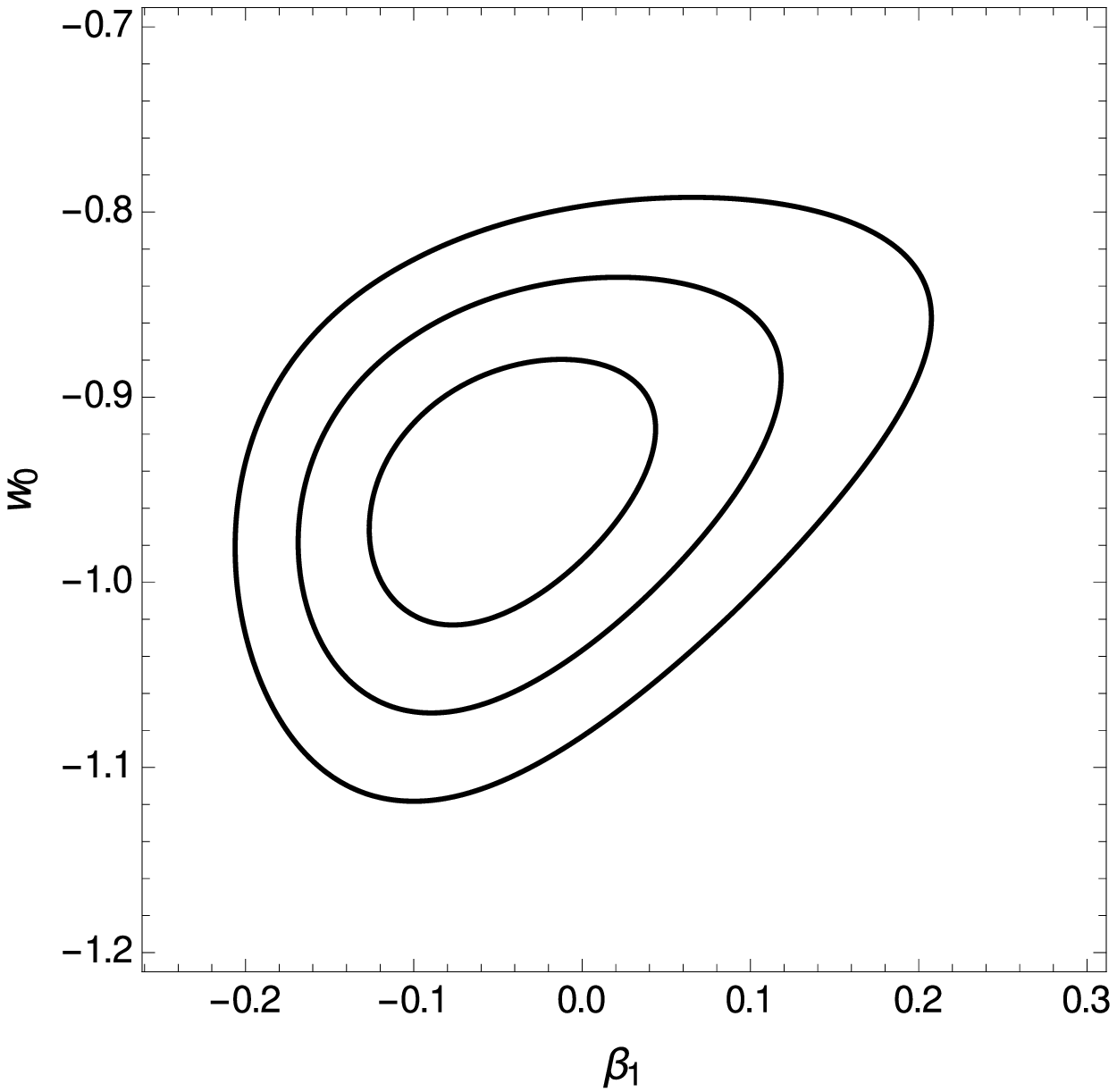}   \\
    \includegraphics[width=.35\textwidth]{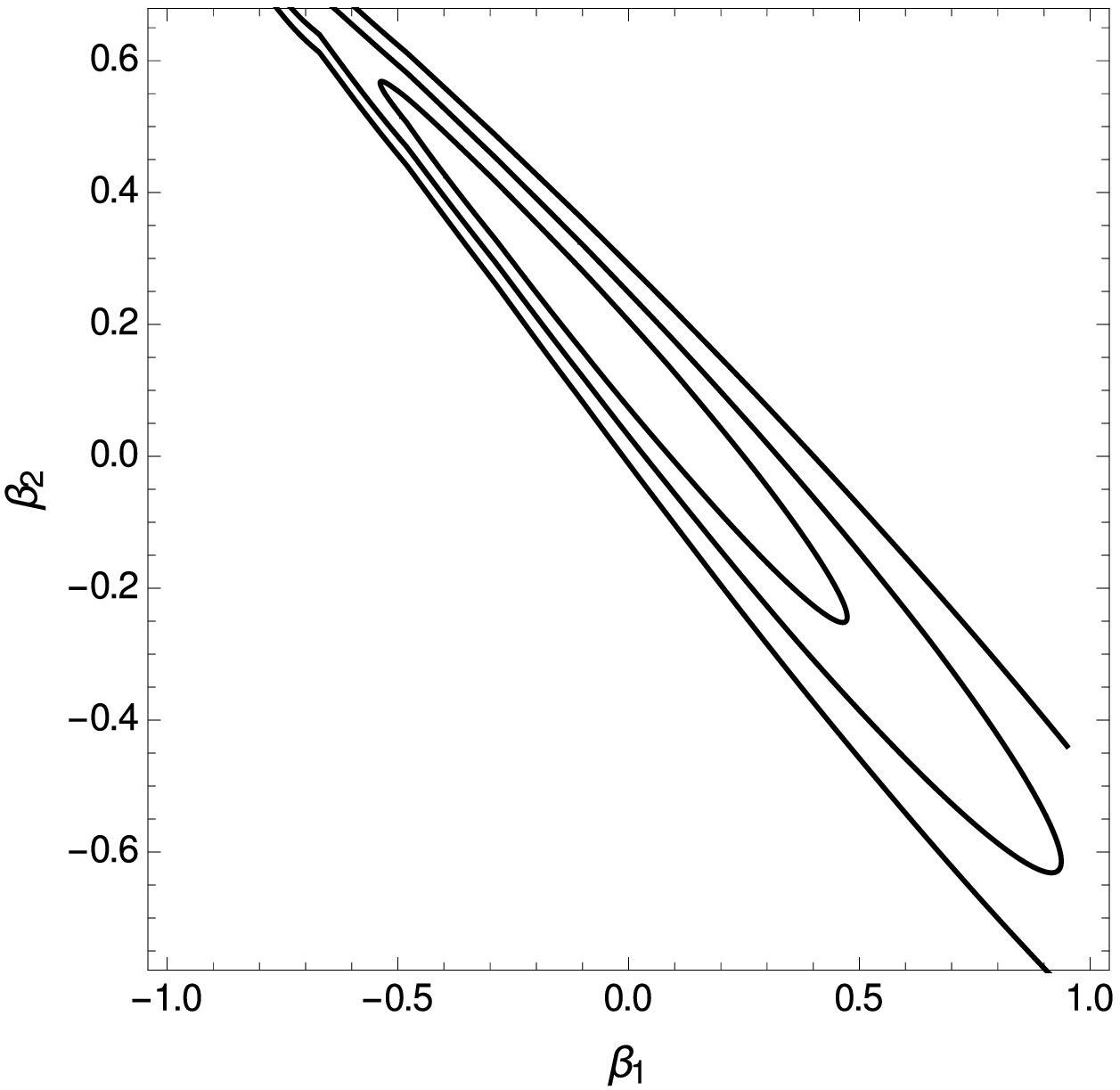} &
  \includegraphics[width=.48\textwidth ]{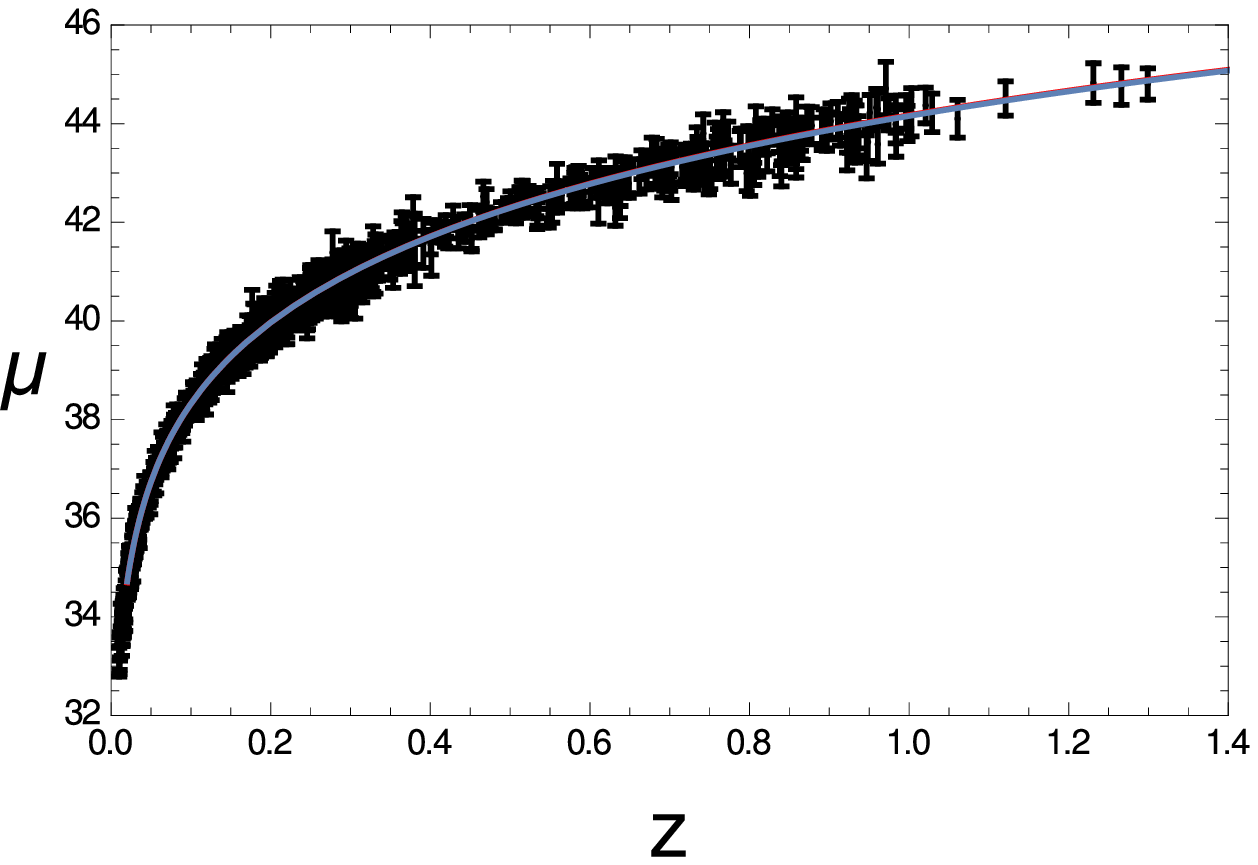}
  \end{tabular}
    \caption{Background dynamics:Results of joint analysis by using $SNIa$ and $H(z)$ data. Here we have used $r_0=1/3$.}
    \label{fig:fig1}
\end{figure}

The results of the  joint analysis are presented in Fig.\ref{fig:fig1}.
The upper left-panel shows the constraint on the  $\beta_2$-$w_0$ plane after marginalization in the parameter $\beta_1$,
the upper right-panel  represents the $\beta_1$-$w_0$ plane after marginalization in $\beta_2$,
and the lower left-panel
shows  the $\beta_1$-$\beta_2$ plane after of the marginalization in the EoS parameter $w_0$.
Also,  in the lower right-panel, we show the plot of the distance
luminosity in terms of the redshift $z$ for the best-fit values using the joint data, JLA SNIa +$H(z)$.
In all the figures the lines represent the contours of the region corresponding to 1$\sigma$, 2$\sigma$ and 3$\sigma$.

From the background dynamics
we find that the values of the best-fit for the parameters $\beta_1$, $\beta_2$, and $w_0$
are $\beta_1=-0.05^{+0.05+0.08+0.10}_{-0.05-0.07-0.09}$, $\beta_2=0.18^{+0.04+0.06+0.08}_{-0.04-0.06-0.08}$ and
$w_0=-0.95^{+0.05+0.06+0.08}_{-0.05-0.07-0.09}$, respectively.

It is interesting to note that  the parameter $\beta_2 >\beta_1$
and that the present value for the EoS parameter $w_0$ is
well supported by observational data \cite{C4,C5}.
In spite that the  model is in well-agreement with
background data ($\chi^2_{\nu} \sim 1.12$), we will see in the
next section, that the general model, together with its best-fit model,
presents instabilities at perturbative level doing inviable the
processes of structure formation and CMB anisotropies. In order to
avoid the instabilities we can mention that it will be possible
only reconstruct the model from an analysis of the perturbative
dynamics. However, the reconstruction will have implications in
the best-fit background model.

In the next section we will study the
linear perturbations and instabilities for our interacting Ricci DE model.

\section{Linear perturbations  and instabilities:  interacting Ricci dark energy model}

In order to study the linear
perturbations and the instabilities in our concrete example, we find that the perturbation, at linear order, of the interaction term
$\delta Q^c$ can be written as

\begin{eqnarray}
\label{finaldeltaQ}
\delta Q^c=-{\epsilon_m}\left[aHv_1\delta^{c \prime}+g_3\delta^c+g_4S^c\right],
\end{eqnarray}
where the functions $g_3$ and $g_4$ are given by
$$
g_3\equiv \frac{3H}{1+r+w}\left[q_mw-r(\beta_1-\beta_2)\right]-3H\beta_2,\,\,
\mbox{and}\,\,g_4\equiv 3H\frac{r}{1+r+w}(\beta_1-\beta_2)+3Hq_m,
$$
respectively. Here we have used Eqs.(\ref{Density}) and (\ref{interaction}).

Now replacing Eq.(\ref{finaldeltaQ}) in Eq.(\ref{SppQ}), we obtain that  the equation for $S^c$ becomes
$$
S^{c\,\prime \prime}+\left(1+y_3+m\right)\frac{S^{c \,\prime}}{a}+\left(ay'_3+y_3m+\frac{k^2y_1}{a^2H^2}\right)\frac{S^c}{a^2}=
-\left[av'_3+v_3m+\frac{k^2v_1}{a^2H^2}\left(\frac{\hat{c}^2_s}{q_m}-y_1\right)\right]\frac{\delta^c}{a^2}
$$
\begin{eqnarray}\label{sfinal}
- \left(v_3-v_1m-av'_1-v_1 \right)\frac{\delta^{c \,\prime}}{a} +v_1\delta^{c \,\prime
\prime},
\end{eqnarray}
where
$$ y_3=3y_1(1-q_m)-\frac{3\,r(\beta_1-\beta_2)}{1+r+w} ,\,\,\mbox{and}\,\,
v_3=3v_1\left[y_1(q_m-1)-1-\hat{c}^2_s\right]
+\frac{3(\beta_2+r \beta_1+w\beta_2 )}{1+r+w} \,.
$$
Combining solutions of Eqs.(\ref{deltafinal1}) and (\ref{sfinal})  for our specific model we find that the matter density perturbation
comoving to dark matter, results in
\begin{eqnarray}
\label{deltamcm}
\delta^{c_m}_m =\frac{1+r}{1+r+w}\delta^c-S^c-3H^2(q_m-1)\frac{a^2}{k^2}
\left(aS^{c \prime}+v_3\delta^c+y_3S^c-av_1\delta^{c \prime}\right),
\end{eqnarray}
and for the holographic Ricci dark energy density perturbation comoving to dark matter we get
\begin{eqnarray}
\label{deltaxcm}
\delta^{c_m}_H =\frac{w}{1+r+w}\delta^c+S^c+
3H^2(1+r+w)\frac{a^2}{k^2}\left(aS^{c \prime}+v_3\delta^c+y_3S^c-av_1\delta^{c
\prime}\right).
\end{eqnarray}
Here we have used  Eq.(\ref{dpcx1}).

In order to know the evolution of the
perturbations $\delta^{c_m}_m $, it is necessary solve the  system
of equations given by (\ref{deltafinal1})--(\ref{sfinal}).
However, first we shall analyze the initial conditions of this
perturbations considering the high-redshift limit.

.


\subsection{Analysis at High-redshifts and Comparing with $\Lambda$CDM}
In the following we will analyze our results
in high-redshift limit, in which  $z\gg 1$ (or equivalently $a\ll1$). At
this limit we obtain that
 the expressions for the quantity $q_m$, given by Eq.(\ref{qm}),  the rate $r$, and  the EoS parameter
 $w$ from Eq.(\ref{rwQ}) become
\begin{eqnarray}
\label{hzlimit}
q_m\rightarrow\frac{C\beta_2-D\beta_1}{C-D},\,\,\,\,\,r\rightarrow-\frac{D}{C},\,\,\mbox{and}
\;\;\,w\rightarrow-\frac{D}{3C}-\frac{r_0}{3}+w_0,\,\,\,\,\,(a\ll1)
\end{eqnarray}
where $C$ and $D$ are given by Eq.(\ref{CDQ}).   In  particular, in the non-interacting case, i.e.,
 $\beta_1=\beta_2=0$ where $C=1/3$ and $D=w_0-\frac{r_0}{3}$, we find that  at high- redshift limit
$q_m\rightarrow 0,\,\,\,\,\,r\rightarrow r_0-3w_0,$ and $\,w\rightarrow 0$.

\begin{figure}[htb]
    \includegraphics[]{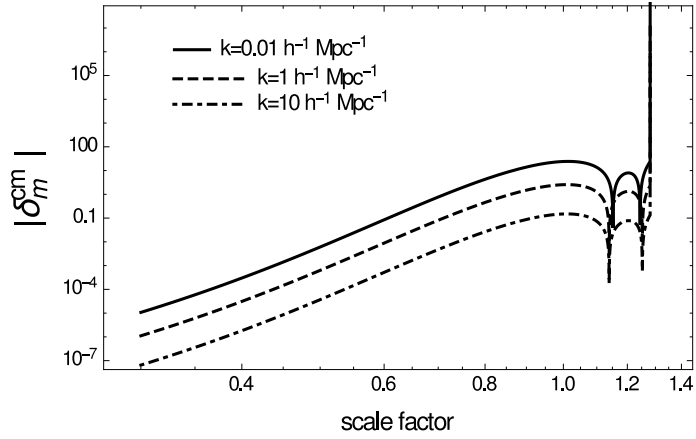}
    \caption{The evolution of the perturbation $\delta^{cm}_m$ versus the scale factor $a$,
    for three different scales $k$. Here we have considered the best fit values of our parameters in which
    $\beta_1=-0.05$, $\beta_2=0.18$ and $w_0=-0.95$, respectively.   }  \label{fig4}
\end{figure}

On the other hand, we obtain that the Eq.(\ref{deltafinal1}) at high-redshift reduces to
\begin{eqnarray}
\label{deltahz}
\delta^{c \prime \prime}+c_1\frac{\delta^{c\prime}}{a}+c_2\frac{\delta^{c}}{a^2}=0,\;\;\,\,\,\,(a\ll1),
\end{eqnarray}
where
$c_1\approx \frac{3}{2}$ and $c_2\approx -\frac{3}{2}$.
Here we have considered  the value $r_0\approx \frac{1}{3}$ and the values of the best-fit:
$\beta_1=-0.05$, $\beta_2=0.18$ and $w_0=-0.95$.

In particular for the non-interacting case ($\beta_1=\beta_2=0$), the Eq.(\ref{deltafinal1})  approaches
 exactly to the Einstein-de Sitter limit
\begin{eqnarray}
\label{EdSl}
\delta^{c \prime \prime}+\frac{3}{2a}\delta^{c \prime}-\frac{3}{2a^2}\delta^{c}=0, \,\,\,\,\,(a\ll1).
\end{eqnarray}
Here we observe that Eqs.(\ref{deltahz}) and (\ref{EdSl}) are very similar, since the constants
$c_1\approx- c_2\approx  2/3$ in Eq.(\ref{deltahz}).

Now  considering   Eq.(\ref{hzlimit}),  we find that the Eq.(\ref{sfinal}) at high-redshift becomes
\begin{eqnarray}
\label{Schzc}
S^{c \prime \prime}+c_3\frac{S^{c\prime}}{a}+c_4\frac{S^c}{a^2}+c_5 \delta^{c\prime \prime}
+c_6\frac{\delta^{c \prime}}{a}+c_7\frac{\delta^c}{a^2}=0, \,\,\,\,\,(a\ll1),
\end{eqnarray}
where
$c_3\approx \frac{3}{2}$ and $c_4\approx c_5\approx c_6\approx c_7\approx 0$. As
before we have used the best -fit values.

For the non-interacting limit ($\beta_1=\beta_2=0$), we obtain
\begin{eqnarray}
\label{Schz}
S^{c \prime \prime}+\frac{3}{2a}S^{c \prime}=0,\,\,\,\,\,(a\ll1).
\end{eqnarray}
Again we note that Eqs.(\ref{Schzc}) and (\ref{Schz}) are very similar for $S^{c \prime}$, since the constants
$c_4\approx c_5\approx c_6\approx c_7\approx 0$.
Also we observe that the solution of $S^c$ from Eq.(\ref{Schzc}) (or Eq. (\ref{Schz}))
 has two modes; one decaying solution (nonphysical) and  the  solution; $S^c$=constant. This last solution suggests
 that we could consider the  adiabatic condition or equivalently $S^c\approx 0$ for the coupled system (see e.g., Ref.\cite{Hipolito2010}).

In Fig.\ref{fig4} we show the evolution of $\delta^{cm}_m$ as a function of the
scale factor $a$ for three different scales $k$. In order to write down values for the perturbation $\delta^{cm}_m$
and the scale factor,
 we solve numerically  the Eqs. (\ref{deltafinal1}) and (\ref{sfinal}) by using the best fit
values found in the previous section and adiabatic initial conditions.
 We observe  that the behavior of $\delta^{cm}_m=\delta^{cm}_m(a)$ on different scales $k$ is similar.
We also note that the perturbations  slowly increase until approximately the value $a=1$
and then they start to oscillate before diverge.
This instability occurs  when the  EoS parameter $w$ crosses the value $w=-1$ in its evolution, known as the phantom crossing , and then the perturbation
collapses at this time and, in particular, at future values for the scale factor  $a>1$.

It is interesting to compare our results
 for the matter perturbations of our holographic model  with the behavior of the matter
perturbations in $\Lambda$CDM model. Here we mention that the $\Lambda$CDM model
 can be obtained as a specific  case of  Eq.(\ref{deltafinal1}), since  in this model
 the pressure
   $p=p_{\Lambda}=-\epsilon_{\Lambda}$=constant and then  we have
\begin{eqnarray}
\epsilon &=& \epsilon_m+\epsilon_{\Lambda},\,\,\,\;\;\;\frac{p}{\epsilon}=
-\frac{\epsilon_{\Lambda}}{\epsilon_{\Lambda}+\epsilon_m}=-\frac{1}{1+r},\nonumber\\
\mbox{and}\,\,\,\,\,r &=& r_0a^{-3}\,\,\,\,\,(\Lambda CDM).\label{RLC}
\end{eqnarray}
In this form, Eq.(\ref{deltafinal1}) reduces to
\begin{equation}
\delta^{c \prime \prime}+\left[\frac{3}{2}-\frac{15}{2}
\frac{p}{\epsilon}\right]\frac{\delta^{c \prime}}{a}-
\left[\frac{3}{2}+12\frac{p}{\epsilon}-\frac{9}{2}
\frac{p^2}{\epsilon^2}\right]\frac{\delta^c}{a^2}=0 \,\,\,(\Lambda CDM).\label{LCDM}
\end{equation}
Now considering that the perturbations $\delta \epsilon^c=\delta \epsilon_m^c$
since $\delta \epsilon_\Lambda=0$, we find that the relation between $\delta^c$
and $\delta^c_m$ can be written as
\begin{eqnarray}
\delta^c=\frac{\delta \epsilon^c}{\epsilon}=\frac{\delta \epsilon_m^c}{\epsilon_{\Lambda}+\epsilon_m}=\delta_m^c\frac{r}{1+r}\,\,\,\,\,(\Lambda CDM),
\end{eqnarray}
and replacing this relation in Eq.(\ref{LCDM}) we obtain the standard  expression  for the matter energy
perturbation given in $\Lambda CDM$ model
\begin{eqnarray}
\delta_m^{c \prime \prime}+\frac{3}{2}\left(\frac{2+r}{1+r}\right)\frac{\delta_m^{c \prime}}{a}-
\frac{3}{2}\left(\frac{r}{1+r}\right)\frac{\delta_m^{c}}{a^2}=0,\,\,\,\,\,(\Lambda
CDM)
\end{eqnarray}
with the ratio $r=r(a)$ given by Eq.(\ref{RLC}).

\section{Avoiding instabilities and an appropriate model  }
In Sect. IV it was demonstrated that the perturbative dynamics of some dark energy models
with a dynamical EoS  parameter $w$  presents instabilities. The instabilities in the structure formation
at linear regime are strongly related to the condition $1+w=0$ at finite time.
In our analysis,
the quantity $1+w$ appears in the coefficient $\tilde{A}$ of  Eq.(\ref{AA}) and  is hidden in the  $\delta p^{c_x}$ term of function $\tilde{B}$
(see Eq. (\ref{deltapcxQ})).  These are the sources of instabilities in
the perturbative dynamics for any dark energy model with a dynamical EoS parameter $w$ with crossing phantom.
 As we noted, such perturbations become very large when the parameter $w$ approaches to
 $w\sim-1$. We also noted that this feature does not depend on the interaction term $Q$, since the
 quantity $1+w$ does not appear in the perturbation of the interaction term $\delta Q^c$.


In the following  we will analyze the conditions for which the quantity $1+w$ becomes zero
and study how to avoid this situation in our  specific model, for any finite time.

From Eq.(\ref{rwQ}) we obtain that the quantity $1+w$  can be written as
\begin{eqnarray}
\label{1plusw} 1+w = \left[\frac{D
r_0-Cr_0\left(3w_0-r_0+3\right)+\left(3w_0-r_0+3\right)
\left(Cr_0+D\right)a^{-3D}}{3\left(Cr_0+D\right)a^{-3D}
-3Cr_0}\right] \,.
\end{eqnarray}

In order to avoid the instabilities
 we can consider
that the quantity $1+w=0$ occurs  for a determined value $a_i$ of
the scale factor in  Eq.(\ref{1plusw}), yielding
\begin{eqnarray}
\left[D
r_0-Cr_0\left(3w_0-r_0+3\right)+\left(3w_0-r_0+3\right)\left(Cr_0+D\right)a^{-3D}\right]|_{a=a_i}=0
\,,
\end{eqnarray}
in which for $a_i$ we get
\begin{eqnarray}
a_i^{-3D}=\frac{Cr_0\left(3w_0-r_0+3\right)-D
r_0}{\left(3w_0-r_0+3\right)\left(Cr_0+D\right)} \,,\label{ai}
\end{eqnarray}
where $-3D=r_0-3w_0-\beta_2>0$.
In order to evade the
singularities in any finite time, we may considered that the scale factor  $a_i \rightarrow \infty$
at a finite time in the future. In this form
, from Eq.(\ref{ai}),
we obtain that the
condition $a_i \rightarrow \infty$ is satisfied in two cases; i) $Cr_0+D=0$ or ii)  $3w_0-r_0=-3$.
Analyzing separately for  both cases we have that:

For the case in which   $Cr_0+D=0$ (or equivalently
$r_0=-(\beta_2+w_0)/\beta_1$) we get
\begin{eqnarray}
\frac{w}{1+r}=\frac{1}{3}\frac{D-C\left(3w_0-r_0\right)}{(D-C)} =
\frac{w_0}{1+r_0}.
\end{eqnarray}
Now, replacing in Eq.(\ref{friedmann}) we find that the Hubble rate in
terms of the scale factor $a$ becomes
\begin{eqnarray}
\label{hnonreal}
H=H_0a^{-\frac{3}{2}\left(1+\frac{w_0}{1+r_0}\right)}
\,\,\,\,\;\;\;\mbox{in which}\;\;\;\;\,\,\,\,a(t)\propto
t^{\frac{2}{3(1+w_0/(1+r_0))}}.
\end{eqnarray}
Here, we note
that this Hubble rate does not depend on the parameters which characterize the interaction term.
We also note that this expression for the Hubble rate describes an universe  without
an acceleration phase in the case $1+r_0>-w_0$, then the model is
disproved from observations. The acceleration phase occurs when
$1+r_0<-w_0$, however this condition corresponds to phantom model,
since $r_0>0$ then $w_0<-1$ and the instabilities take place at values for the scale factor such that  $a_i<1$  (past
time), then
we discarded this phantom model. In this
way, the first condition $Cr_0+D=0$ is not a suitable condition to avoid the instabilities.

Now let us analyze the second condition $3w_0-r_0=-3$. From this
condition  we note that the values $r_0$ and $w_0$  cannot be
taken  independently. Also, we note that  as $r_0>0$ then the
parameter $w_0>-1$. Considering the second condition we get that
the parameter which characterizes the Ricci DE, $c^2$, becomes $c^2=2(r_0-3w_0+1)^{-1}=1/2$, independently of
the values of $r_0$ and $w_0$. This result for the parameter
$c^2=1/2$ of the Ricci DE coincides with the obtained in
Refs.\cite{DelCampo2013,V0} from the analysis of the non-interacting case, where the instabilities in
the non-adiabatic perturbations were considered.

 As before we obtain  that the quantity
$1+w$ as a function of the scale factor becomes
\begin{eqnarray}
1+w=1+\frac{D r_0+3Cr_0-3\left(Cr_0+D\right)a^{-3D}}{3\left(Cr_0+D\right)a^{-3D} + 3(D-C)r_0}=
\frac{4D r_0}{3\left(Cr_0+D\right)a^{-3D} + 3(D-C)r_0}\,,
\end{eqnarray}
and from  Eq.(\ref{friedmann}), the solution
for the  Hubble rate becomes
\begin{eqnarray}
H=H_0a^{-3/2}\left[\frac{\left(Cr_0+D\right)a^{-3D}+\left(D-C\right)r_0}{D(1+r_0)}\right]^{-1/2D}
\,,\label{Hg}
\end{eqnarray}
or equivalently
\begin{eqnarray}
H=H_0a^{-3/2}\left[\left(1+\frac{3(\beta_1-\beta_2)+4}{3(\beta_2-1)}
\Omega_{m0}\right)\,a^{-3(\beta_2-1)}-\frac{3(\beta_1-\beta_2)+4}{3(\beta_2-1)}
\Omega_{m0}\right]^{-\frac{1}{2(\beta_2-1)}}.\label{Hg2}
\end{eqnarray}
Here we have used that $1=\Omega_{m0}+\Omega_{H0}$, then
$\Omega_{m0}=3(1+w_0)/[1+3(1+w_0)]$.
\begin{figure}[htb]
\includegraphics[width=.35\textwidth]{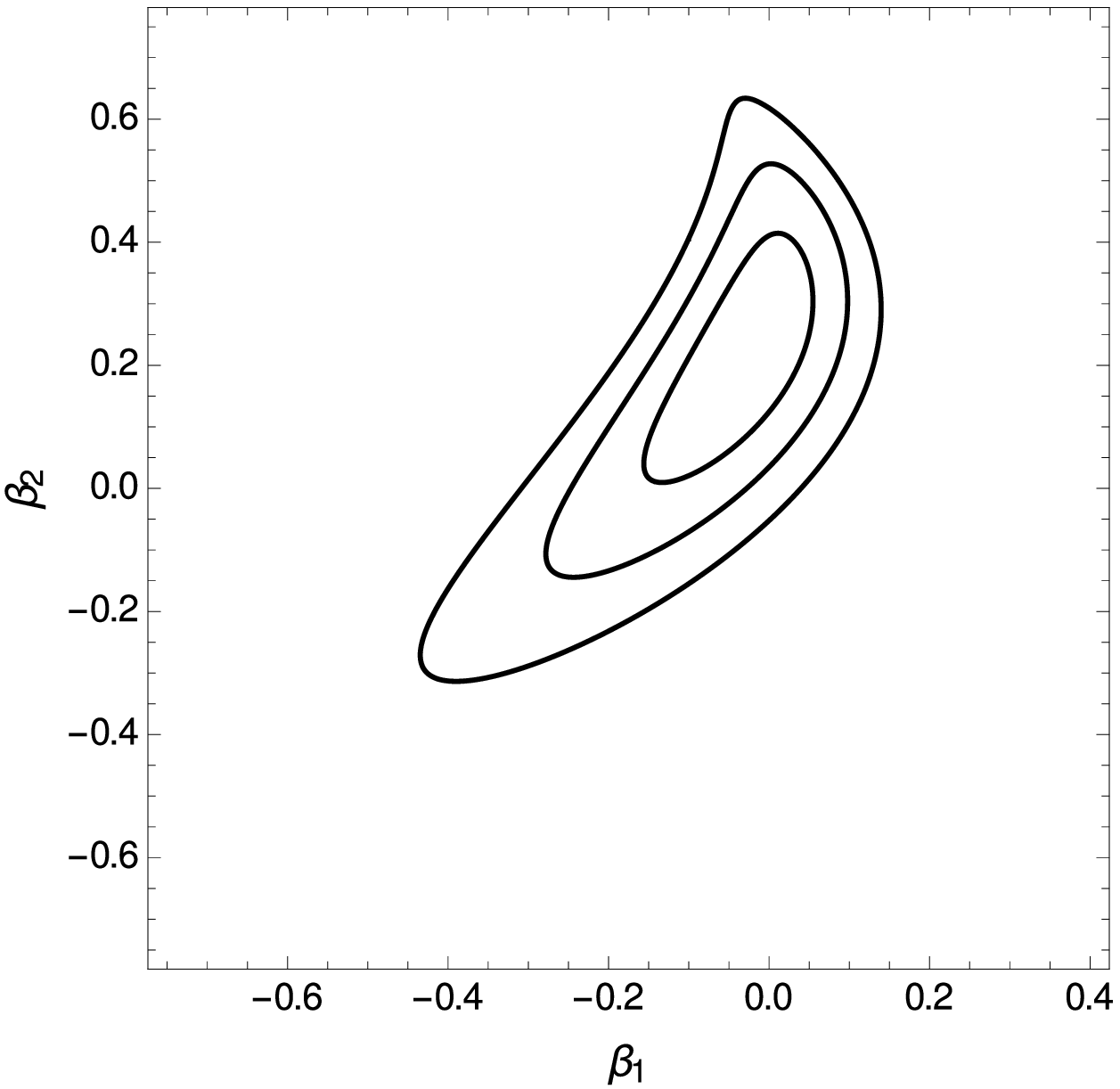}
\includegraphics[width=.35\textwidth]{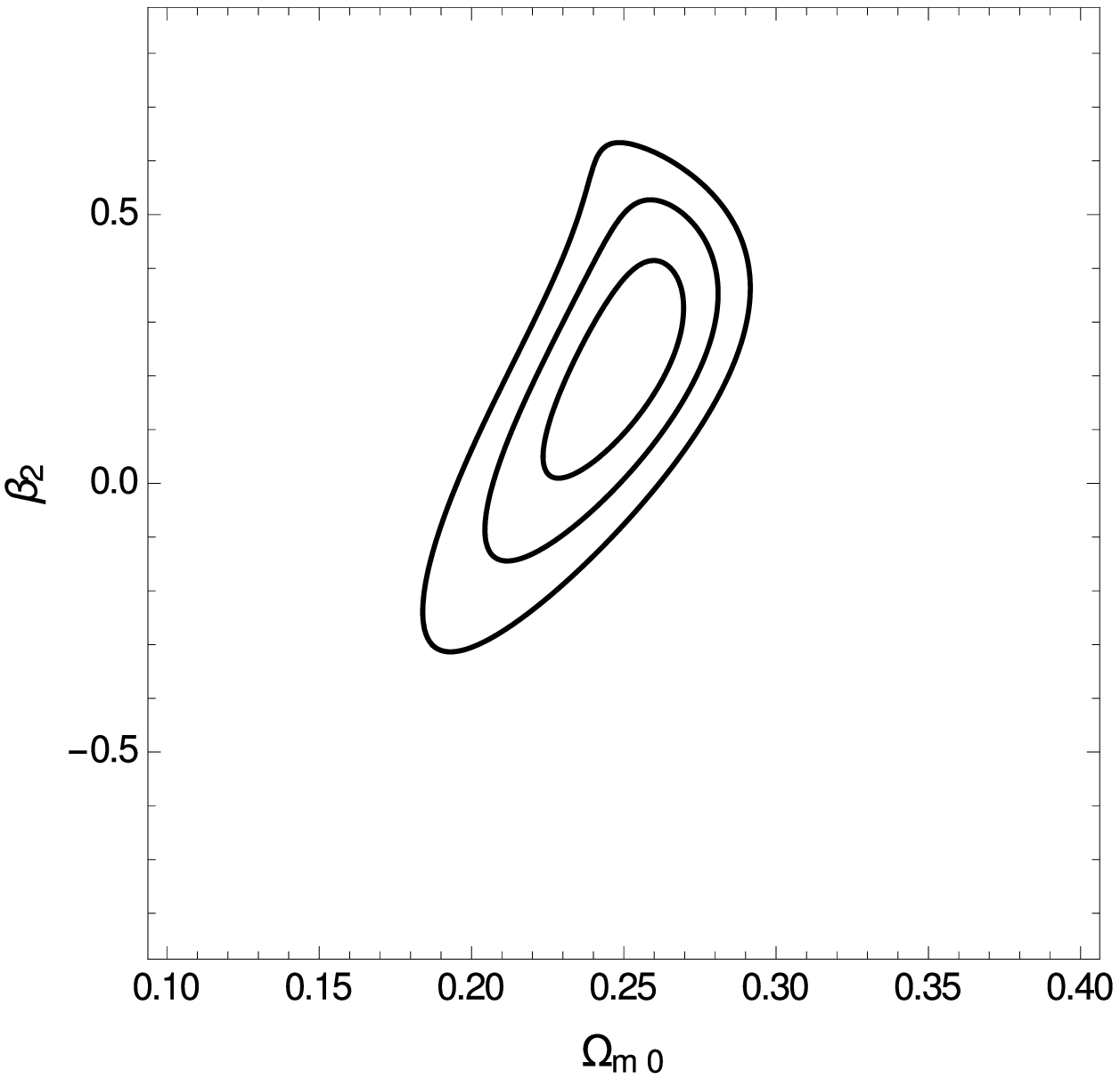}
    \caption{Model without instabilities: Results of joint analysis by using SNIa and $H(z)$ data considering the second condition. Here we have considered Eq.(\ref{Hg2}) and
     $\Omega_{m0}=0.25$. } \label{fig5}
\end{figure}

As before, we perform the same analysis of  section V, considering the
SNIa and $H(z)$ data sets.
 In Fig.(\ref{fig5}) we show the constraints on the $\beta_1-\beta_2$
 plane after marginalization in the parameter $\Omega_{m0}$ and the constraints on the $\beta_2-\Omega_{m0}$
 plane after marginalization in the parameter $\beta_1$,
 considering  Eq.(\ref{Hg2}). Again the lines represent the contours of the 1$\sigma$, 2$\sigma$ and 3$\sigma$ regions, respectively.
We find that the values of the best-fit for the parameters $\beta_1$,
$\beta_2$ and $\Omega_{m0}$ become
$\beta_1=-0.03^{+0.04+0.07+0.08}_{-0.10-0.14-0.17}$,
$\beta_2=0.22^{+0.09+0.14+0.18}_{-0.10-0.14-0.17}$, and
$\Omega_{m0}=0.25^{+0.01+0.03+0.04}_{-0.01-0.02-0.03}$,
respectively ($\chi^2_{\nu} \sim 1.17$). We note a drastic change in the values of the parameters
 in order to avoid the appearance of instabilities in the perturbative
dynamics. We also note that form the best-fit, $\beta_1>\beta_2$, then
the direction of energy transfer is from DE to DM, since $q_m>0$, or equivalently $Q>0$ (figure not shown). 
In particular, the value $\beta_1$ changes from
$\beta_1=-0.05$ to $\beta_1=-0.03$, which represents an increase about of 40
percent, and for $\beta_2$ an increase about of 22$\%$.

\begin{figure}[htb]
\includegraphics[width=.5\textwidth]{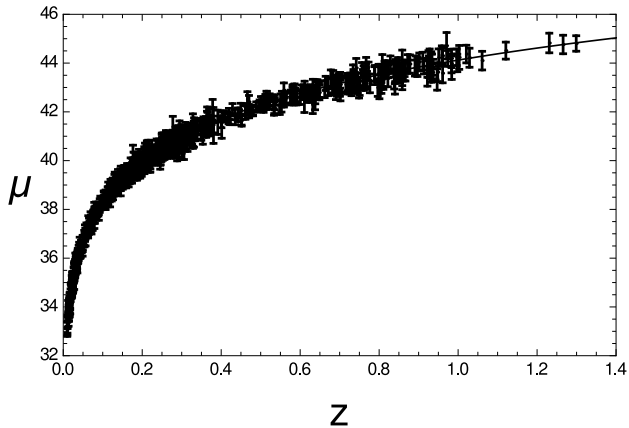}
    \caption{Model without instabilities: Plot of the
luminosity distance $\mu$ versus the redshift $z$ for the best-fit values
in contrast with JLA SNIa data and considering  the second condition
i.e., $3w_0-r_0=-3$. Here we have used the values $\beta_1=-0.03$,
$\beta_2=0.22$. }\label{anfig1}
\end{figure}

In Fig.(\ref{anfig1}) we show the plot of the luminosity distance
$\mu$ versus the redshift $z$ for the best-fit values
in contrast with JLA SNIa data and considering  the second condition
$3w_0-r_0=-3$. Here we have used the values $\beta_1=-0.03$,
$\beta_2=0.22$ and $\Omega_{m0}=0.25$ (best-fit values avoiding instabilities).
 We note that the solution of the Hubble rate, given by Eq.(\ref{Hg2}),
presents an accelerate phase and is well supported by the
observational data.

 On the other hand, it is interesting to compare our
 model with the holographic Ricci DE model without DE-DM
interaction, considering the  \textit {reduced chi square} $\chi^2_\nu$  statistical function 
with the same data set.  In this form, we could  determine which
type of model best fits to the observation data, observing     that model presents the  lowest  value of  $\chi_\nu^2$. In the case
without DE-DM interaction, the Hubble rate for the model without
instabilities is given by \cite{DelCampo2013}
\begin{eqnarray}
H=H_0\left[\left(1-\frac{4}{3}
\Omega_{m0}\right)\,+\frac{4}{3}
\Omega_{m0}\,a^{-3}\right]^{\frac{1}{2}}.\label{Hg2}
\end{eqnarray}
 Here, we note that  we have only one free parameter, the  matter density $\Omega_{m0}$. By
performing joint tests using JLA SNIa  and $H(z)$ data set we
obtain that $\Omega_{m0}=0.22^{+0.01+0.03+0.04}_{-0.01-0.03-0.04}$ with $\chi^2_\nu \sim 1.18$.
However, we observe that this goodness-of-fit is slightly higher than
the case with interaction, since we have found the value $\chi^2
_\nu \sim 1.17$. Unfortunately, because of the similitude in the
values of $\chi^2_\nu$ obtained in both models, we cannot directly
deduce which model realizes better the observational data.
Strictly, our model presents a lower value in relation to the non-interacting model, then we may conclude
that the most general case with energy transfer between DE-DM
 is well suited for to study and is well confronted with  the
observational data.

\begin{figure}[htb]
    \includegraphics[]{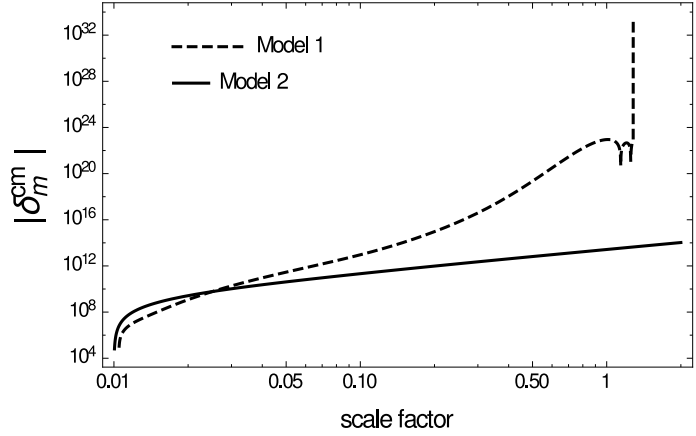}
    \caption{The evolution of the perturbation $\delta^{cm}_m$ as function of the scale factor $a$,
    for the  scale $k=0.05 h^{-1} Mpc^{-1}$. Here we have used the values $\beta_1=-0.05$, $\beta_2=0.18$ and $w_0=-0.95$ for the model 1,
    and $\beta_1=-0.03$, $\beta_2=0.22$ and $\Omega_{m0}=0.25$ for the model 2.  }\label{fig7}
\end{figure}

\begin{figure}[htb]
    \includegraphics[]{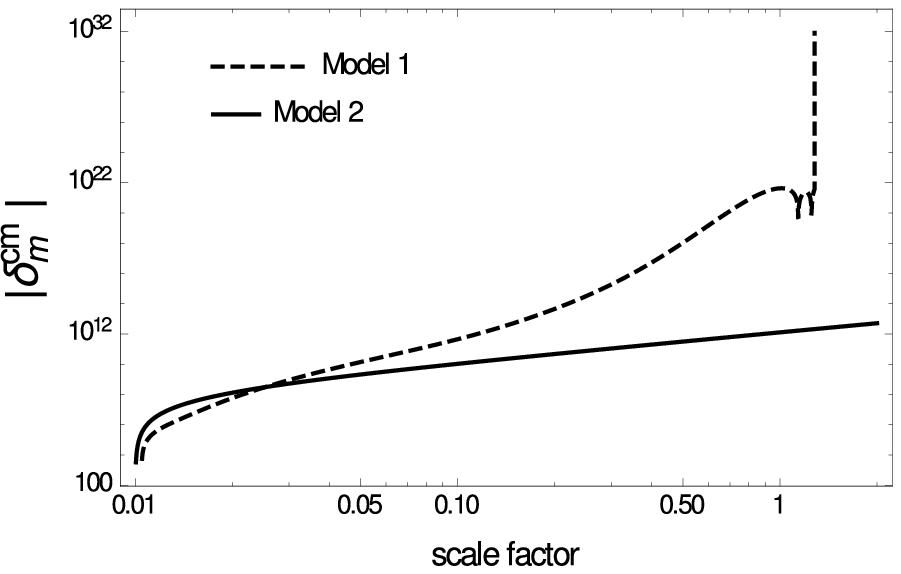}
    \caption{The evolution of the perturbation $\delta^{cm}_m$ as function of the scale factor $a$,
    for the  scale $k=1.5 h^{-1} Mpc^{-1}$. Here we have used the values $\beta_1=-0.05$, $\beta_2=0.18$ and $w_0=-0.95$ for the model 1,
    and $\beta_1=-0.03$, $\beta_2=0.22$ and $\Omega_{m0}=0.25$ for the model 2.}\label{fig8}
\end{figure}

In Figs.(\ref{fig7}) and (\ref{fig8}) we show the evolution of the perturbation $\delta^{cm}_m$
as a function of the scale factor $a$ for two different scales $k$: $k=0.05 h^{-1} Mpc^{-1}$ and $k=1.5 h^{-1} Mpc^{-1}$, respectively.
 Here,
we have denoted as the Model 1 the model with instabilities and
the Model 2  as the model without instabilities considering the   second condition $3w_0-r_0=-3$ .
 As before, we find numerically the solutions for the coupled system Eqs. (\ref{deltafinal1}) and (\ref{sfinal}) by
considering  the best fit values found from the data analysis for
both models, see Figs. (\ref{fig:fig1}) and (\ref{fig5}).   Here
we have used the values $\beta_1=-0.05$, $\beta_2=0.18$ and $w_0=-0.95$ for
the model 1, and $\beta_1=-0.03$, $\beta_2=0.22$ and
$\Omega_{m0}=0.25$ for the model 2.  We observe that by considering
the   second condition $3w_0-r_0=-3$ (model 2), we can evade the
instabilities in the structure formation and then we obtain an
appropriate model for the dark sector from the perturbative
analysis.

\section{Conclusions}

In this paper we have analyzed  an interacting model of dark
energy and dark matter in order to describe of late cosmic
acceleration of the universe. Under a general formalism  we have
described the perturbative dynamics for these two interacting
fluids. In this general analysis we have considered the timelike part
of the balance equation, the momentum balance and the momentum
transfer $Q^{\alpha}$ associated to the interaction term $Q$. From
the functions   of  contrast for dark  matter and dark energy we
have studied the perturbative dynamics considering a
gauge-invariant treatment in comoving gauge.
 On the other hand, we have obtained  the total non-adiabatic
and dark energy pressure perturbations and we found the relation
between these gauge-invariant quantities. Also, we
have identified  the sources of instabilities in DE models with a
dynamical EoS parameter that presents a  phantom crossing. As a
concrete example we have considered an interaction term $Q$
between the holographic Ricci-DE and DM. Here we have studied that
the interaction term $Q$, depends on the energy densities of both
components multiplied by a quantity with units of the inverse of
time (proportional to the Hubble parameter). From the background
equations we  have obtained the constraints on the parameters
characterizing the interaction by considering the observational
analysis from the SNIa and $H(z)$ tests. Here from the background
dynamics we have found   that the best-fit values for the
parameters of the interaction are
$\beta_1=-0.05^{+0.05+0.08+0.10}_{-0.05-0.07-0.09}$,
$\beta_2=0.18^{+0.04+0.06+0.08}_{-0.04-0.06-0.08}$, and for the
EOS parameter $w_0=-0.95^{+0.05+0.06+0.08}_{-0.05-0.07-0.09}$. In
our perturbative analysis we have found that, in
this best-fit model, the instabilities appear at the moment when
the EoS parameter $w$ crosses the value $w\sim -1$ and we have
noted that this feature does not depend on the interaction term
$Q$. In order to avoid these instabilities in the perturbative
analysis and develop an appropriate model  for any finite time, we
have obtained a specific  value of the scale factor denoted as
$a_i$. From this value of $a_i$ we have obtained  two independent
conditions to avoid the instabilities in our specific model,
namely: i) $Cr_0=-D$ and ii) $r_0=3(w_0+1)$. Considering the first
condition we have found that if $w_0<1+r_0$, the model is
disproved from observations, since under this requirement the
model does not present an accelerate scenario. Otherwise if
$w_0>1+r_0$, we have obtained  an accelerate phase, however this
condition corresponds to a phantom model, nothing that the
instabilities take place in the past time. In this way, we have
obtained that the first condition is not suitable. From the second
condition i.e., $r_0=3(w_0+1)$,
 we have
found an accelerate phase of the universe and also corresponds to an
appropriate model  for any finite time. In order to avoid the
instabilities in the perturbative dynamics,  we have noted that this
result agrees with obtained in Ref.\cite{DelCampo2013} and
becomes independently of the interaction term. Moreover we have
obtained that the constraint on the Ricci parameter $c^2=1/2$ is
fixed for the second  condition, since as from background
$c^2=2(r_0-3w_0+1)^{-1}$ and together the second condition
$r_0=3(w_0+1)$, then $c^2=1/2$, independency of the values $r_0$ 
, $w_0$ and the energy transfer rate  $Q$.
 Also, from this condition we have found, in order to have an appropriate model, a new sets of
best-fit values for the interaction parameters given
by $\beta_1=-0.03^{+0.04+0.07+0.08}_{-0.10-0.14-0.17}$,
$\beta_2=0.22^{+0.09+0.14+0.18}_{-0.10-0.14-0.17}$, and
$\Omega_{m0}=0.25^{+0.01+0.03+0.04}_{-0.01-0.02-0.03}$. Here we have observed  a drastic change in the
values of the parameters $\beta_1$ and $\beta_2$ in order to avoid the
singularity from the perturbative dynamics. For the parameter
$\beta_1$ we have found that the increased is the order of  40$\%$
and for $\beta_2$ is the order of  22$\%$.

Finally, we would like to point out that in models that have a phantom crossing, and in particular  
for the holographic models (with and without interaction), 
it is
necessary to be cautious when only the background level  observational
tests are being considered. Here, we have shown that in spite of a
good agreement with data and an adequate background dynamic, this could lead to inviable models at
perturbative level.

\section{Acknowledgments}
R. H. was supported by Comisi\'{o}n Nacional de Ciencias y Tecnolog\'{\i}a of Chile through FONDECYT REGULAR Grant
 N$_{0}$ 1130628 and DI-PUCV  123.724. N. V. was supported by Comisi\'{o}n Nacional de
  Ciencias y Tecnolog\'{\i}a of Chile through FONDECYT  Grant  N$_{0}$ 3150490. WSHR was
supported by Brazilian agencies CAPES (proccess No 99999.007393/2014-08)  at the begining of
this work,  and  FAPES at the end (BPC No 476/2013). WSHR is grateful for the hospitality of the Physics Department of
McGill University where part of this work was developed.



\begin{thebibliography}{}

\bibitem{C1}A. Riess {\it et al.}, Astron. J. {\bf 116}, 1009
(1998).
\bibitem{C2} S. Permutter {\it et al.}, Astron. J. {\bf 517}, 565
(1999).

\bibitem{C3} M. Tegmark {\it et al.}, Astron. J. {\bf 606}, 702
(2004).

\bibitem{C4} P. A. R. Ade {\it et al.},  Astron. Astrophys. A{\bf16},
571 (2014); N.~Aghanim {\it et al.} [Planck Collaboration],
  [arXiv:1507.02704 [astro-ph.CO]].


  \bibitem{C5}  P.~A.~R.~Ade {\it et al.} [Planck Collaboration],
  doi:10.1051/0004-6361/201525941
  arXiv:1502.01591 [astro-ph.CO]; P.~A.~R.~Ade {\it et al.} [Planck Collaboration],
  arXiv:1502.01589 [astro-ph.CO].


 \bibitem{C6}L. Amendola, Phys. Rev. D {\bf 62}, 043511 (2000);
  L. Amendola and C. Quercellini, Phys. Rev. D {68}, 023514 (2003); L.
Amendola, S. Tsujikawa, and M. Sami, Phys. Lett. B {\bf 632}, 155 (2006).

 \bibitem{C7}D. Pavon, W. Zimdahl, Phys. Lett. B {\bf 628}, 206 (2005);
 S. Campo, R. Herrera, D. Pavon, Phys. Rev.D {\bf78}, 021302(R) (2008).
 \bibitem{C8} B. Wang, J. Zang, C.-Y. Lin, E. Abdalla, and S. Micheletti, Nucl. Phys. B {\bf778}, 69
 (2007); S. Cao and N. Liang,  Int. J. Mod. Phys. D {\bf22} (2013).

 \bibitem{C9}Z. K. Guo, N. Ohta, and S. Tsujikawa, Phys. Rev. D {\bf76}, 023508
 (2007); M. Le Delliou, R. Marcondes, G. Lima Neto, and E. Abdalla,  Mon. Not. Roy.
  Astron. Soc. 453, 2 (2015); B.~Wang, E.~Abdalla, F.~Atrio-Barandela and D.~Pavon,
  arXiv:1603.08299 [astro-ph.CO].

 \bibitem{I1}W.~Zimdahl and D.~Pavon,
  Gen.\ Rel.\ Grav.\  {\bf 36}, 1483 (2004); O.~Bertolami, F.~Gil Pedro and M.~Le Delliou,
  Phys.\ Lett.\ B {\bf 654}, 165 (2007).

 \bibitem{I2}W.~Zimdahl,
  Int.\ J.\ Mod.\ Phys.\ D {\bf 14}, 2319 (2005); R.~G.~Cai and A.~Wang,
  JCAP {\bf 0503}, 002 (2005); L.~Amendola, G.~Camargo Campos and R.~Rosenfeld,
  Phys.\ Rev.\ D {\bf 75}, 083506 (2007); G.~Caldera-Cabral, R.~Maartens and L.~A.~Urena-Lopez,
  Phys.\ Rev.\ D {\bf 79}, 063518 (2009).


 \bibitem{AC1} C. G. Park, J. C. Hwang, J. H. Lee, and H. Noh, Phys. Rev. Lett. {\bf103}, 151303 (2009).
\bibitem{AC2} R. Bean and O. Dore, Phys. Rev. D {\bf 69}, 083503 (2004);


\bibitem{AC3}
  J. Weller and A. M. Lewis, Mon. Not. Roy. Astron. Soc. {\bf 346}, 987 (2003);

  
\bibitem{AC4}  G. B. Zhao, J. Q. Xia, M. Li, B. Feng, and Xinmin Zhang, Phys. Rev. D {\bf 72}, 123515 (2005),





 \bibitem{C10}R. Bean, E. Flanagan   and  M. Trodden,  Phys. Rev. D {\bf 78}, 023009
  (2008).

\bibitem{C11} J. Valiviita , E. Majerotto  and R.  Maartens,   J. Cosmol. Astropart. Phys. {\bf07}, 020
(2008).

\bibitem{Kunz:2006wc}
  M.~Kunz and D.~Sapone,
  Phys.\ Rev.\ D {\bf 74}, 123503 (2006).

\bibitem{HP1}
L. Susskind, J. Math. Phys. {\bf36}, 6377 (1995);
J. M. Maldacena, Adv. Theor. Math.
Phys.{ \bf2}, 231 (1998); R. Bousso, Rev. Mod. Phys. {\bf 74},
825 (2000).


\bibitem{HM1} A. G. Cohen, D.B. Kaplan and A.E. Nelson, Phys. Rev. Lett. {\bf 82}, 4971 (1999);

\bibitem{HMM1}  M. Li, Phys. Lett. B {\bf 603}, 1 (2004).

\bibitem{HMM2}  S. Nojiri, S. D. Odintsov,  Gen.Rel.Grav. {\bf 38} 1285 (2006).

\bibitem{HM2}B.~Wang, Y.~g.~Gong and E.~Abdalla,
  Phys.\ Lett.\ B {\bf 624}, 141 (2005); B.~Wang, C.~Y.~Lin and E.~Abdalla,
  Phys.\ Lett.\ B {\bf 637}, 357 (2006); M.~R.~Setare,
  Eur.\ Phys.\ J.\ C {\bf 50}, 991 (2007) ;
  R. C. G. Landim, Int.J.Mod.Phys. D {\bf 25}, No. 4  1650050 (2016).

\bibitem{H0} H.~Kim, H.~W.~Lee and Y.~S.~Myung,
  Phys.\ Lett.\ B {\bf 632}, 605 (2006).
  
\bibitem{HM3}
  W.~Zimdahl and D.~Pavon, Class.\ Quant.\ Grav.\  {\bf 24}, 5461 (2007); Q.~Wu, Y.~Gong, A.~Wang and J.~S.~Alcaniz,
  Phys.\ Lett.\ B {\bf 659}, 34 (2008); J.~Zhang, H.~Liu and X.~Zhang,
  Phys.\ Lett.\ B {\bf 659}, 26 (2008); A.~A.~Sen and D.~Pavon,
  Phys.\ Lett.\ B {\bf 664}, 7 (2008); I.~Duran, D.~Pavon and W.~Zimdahl,
  JCAP {\bf 1007}, 018 (2010);  K.~Das and T.~Sultana,
  Astrophys.\ Space Sci.\  {\bf 361}, no. 2, 53 (2016).

\bibitem{R1}R. Brustein and G. Veneziano, Phys. Rev. Lett. 84, 5695
(2000).
  
\bibitem{V} C. Gao, F. Q. Wu, X. Chen, and Y. G. Shen, Phys. Rev. D
79, 043511 (2009).

  \bibitem{V1}C.~J.~Feng,
  Phys.\ Lett.\ B {\bf 670}, 231 (2008); X.~Zhang,
  Phys.\ Rev.\ D {\bf 79}, 103509 (2009); L.~Xu and Y.~Wang,
  JCAP {\bf 1006}, 002 (2010);  M.~Suwa and T.~Nihei,
  Phys.\ Rev.\ D {\bf 81}, 023519 (2010).

\bibitem{V0} K. Karwan and T. Thitapura, JCAP {\bf 1201}, 017 (2012);  
  
 \bibitem{V2}
 Yuting Wang, Lixin Xu and Yuanxing Gui, Phys. Rev. D {\bf84}, 063513 (2011);
  Chao-Jun Feng and Xin-Zhou Li, Phys. Lett. B{\bf 680}, 355 (2009).

\bibitem{DelCampo2013}  S.~del Campo, J.~C.~Fabris, R.~Herrera and W.~Zimdahl,
  Phys.\ Rev.\ D {\bf 87}, no. 12, 123002 (2013).

\bibitem{bardee} J. Bardeen, Phys. Rev D {\bf 22}, 1882 (1996).

\bibitem{Hipolito2009} W.S.Hip\'olito-Ricaldi, H.E.S. Velten and W. Zimdahl,
JCAP, 06, 016 (2009).

\bibitem{Funo2014} A.Romero Fu\~no, W.S. Hip\'olito-Ricaldi and W. Zimdahl, MNRAS  {\bf 57} no.3, 2958 (2016).

\bibitem{vomMarttens2014} R.F. vom Marttens, W.S. Hip\'olito-Ricaldi and W. Zimdahl, JCAP, 08, 004 (2014).

\bibitem{q1}H. Kodama and M. Sasaki, Prog. Theor. Phys. Suppl. {\bf78}, 1 (1984).

\bibitem{q2}K. A. Malik, D. Wands and C. Ungarelli, Phys. Rev. D {\bf67}, 063516 (2003).

\bibitem{Vargas2012} C. Zu\~niga-Vargas, W.S. Hip\'olito-Ricaldi and W. Zimdahl, JCAP, 04, 032 (2012).

\bibitem{Hipolito2010} W.~S.~Hipolito-Ricaldi, H.~E.~S.~Velten and W.~Zimdahl,
  Phys.\ Rev.\ D {\bf 82}, 063507 (2010).






\bibitem{DelCampo2011} S. del Campo, J.C. Fabris, R. Herrera and W. Zimdahl, Phys.Rev.D 83, 123006 (2011).

\bibitem{Int1}G.~R.~Farrar and P.~J.~E.~Peebles,
  Astrophys.\ J.\  {\bf 604}, 1 (2004);  S.~del Campo, R.~Herrera and D.~Pavon,
  Phys.\ Rev.\ D {\bf 70}, 043540 (2004); R.~Herrera, D.~Pavon and W.~Zimdahl,
  Gen.\ Rel.\ Grav.\  {\bf 36}, 2161 (2004);   S.~del Campo, R.~Herrera, G.~Olivares and D.~Pavon,
  Phys.\ Rev.\ D {\bf 74}, 023501 (2006); 
  Phys.\ Rev.\ D {\bf 75}, 083506 (2007); S.~del Campo, R.~Herrera and D.~Pavon,
  Int.\ J.\ Mod.\ Phys.\ D {\bf 20}, 561 (2011).

  


\bibitem{Re}M.S. Turner, Phys. Rev. D 28, 1243 (1983).


\bibitem{Int2}H.~Wei and R.~G.~Cai,
  Eur.\ Phys.\ J.\ C {\bf 59}, 99 (2009); H.~M.~Sadjadi and M.~Alimohammadi,
  Phys.\ Rev.\ D {\bf 74}, 103007 (2006); A.~Sheykhi,
  Phys.\ Lett.\ B {\bf 681}, 205 (2009); R.~Herrera and N.~Videla,
  Int.\ J.\ Mod.\ Phys.\ D {\bf 23}, no. 08, 1450071 (2014);  M.~Szydłowski, A.~Krawiec, A.~Kurek and M.~Kamionka,
  Eur.\ Phys.\ J.\ C {\bf 75}, no. 99, 5 (2015);  S.~del Campo, R.~Herrera and D.~Pav\'on,
  Phys.\ Rev.\ D {\bf 91}, no. 12, 123539 (2015).




\bibitem{Betoule2014}
M. Betoule {\it et. al.}, 
Astron. Astrophys. {\bf 568}, A22 (2014) . 


\bibitem{Farooq2013} O. Farooq, B. Ratra, Astrophys.J.. 766. L7 (2013).

\bibitem{JLA} http://supernovae.in2p3.fr/sdss\_snls\_jla/ReadMe.html.


\bibitem{Simon2005}
J. Simon, L. Verde and R. Jimenez, , Phys. Rev. D
{\bf 71} (2005) 123001 [arXiv:0412269].

\bibitem{Stern2010}
D. Stern, R. Jimenez, L. Verde, M. Kamionkowski and S.A. Stanford,
 J. Cosmol. Astropart. Phys.
{\bf 02}, 008 (2010)  .



.
















































































\end{thebibliography}
\end{document}